\documentclass[aps,prb,twocolumn,superscriptaddress,floatfix,amsmath,longbibliography]{revtex4-2}
\usepackage{graphicx}
\usepackage{siunitx}
\usepackage{multirow}
\usepackage{amssymb}
\usepackage{hyperref}

\begin{document}
%
\title{Energy states of Rydberg excitons in finite crystals: From weak to strong confinement}
\author{Pavel~A.~Belov}
\author{Florian~Morawetz}
\author{Sjard~Ole~Kr{\"u}ger}
\affiliation{Institut f{\"u}r Physik, Universit{\"a}t Rostock, Albert-Einstein-Stra{\ss}e 23-24, 18059 Rostock, Germany}
\author{Niklas~Scheuler}
\author{Patric~Rommel}
\author{J\"{o}rg~Main}
\affiliation{Institut f{\"u}r Theoretische Physik I, Universit{\"a}t Stuttgart, Pfaffenwaldring 57, 70569 Stuttgart, Germany}
\author{Harald~Giessen}
\affiliation{4th Physics Institute and Research Center SCoPE, Universit{\"a}t Stuttgart, Pfaffenwaldring 57, 70569 Stuttgart, Germany}
\author{Stefan~Scheel}
\affiliation{Institut f{\"u}r Physik, Universit{\"a}t Rostock, Albert-Einstein-Stra{\ss}e 23-24, 18059 Rostock, Germany}

\begin{abstract}
Due to quantum confinement, excitons in finite-sized crystals behave rather differently than in bulk 
materials. We investigate the dependence of energies of Rydberg excitons on the strengths of parabolic as well as rectangular confinement potentials in finite-sized crystals. The evolution of the energy levels of hydrogen-like excitons in the crossover region from weak to strong parabolic confinement is analyzed for different quantum numbers by numerical solution of the two-dimensional Schr\"{o}dinger equation. The energy spectrum of hydrogen-like excitons in Cu$_{2}$O-based rectangular quantum wells is, in turn, obtained numerically from the solution of the three-dimensional Schr\"{o}dinger equation as a function of the quantum well width. Various crossings and avoided crossings of Rydberg energy levels are observed and categorized based on the symmetry properties of the exciton wave function. Particular attention is paid to the two limiting cases of narrow and wide quantum wells attributed to strong and weak confinement, respectively.
The energies obtained with the pure Coulomb interaction are compared with the results originating from the Rytova-Keldysh potential, i.e., by taking into account the dielectric contrast in the quantum well and in the barrier.
\end{abstract}

\maketitle

\section{Introduction}

Hydrogen-like systems are the bedrock of quantum theory, atomic physics as well as solid state physics.
Rotational $SO(3)$ symmetry as well as the hidden $SO(4)$ symmetry properties of the hydrogen 
atom~\cite{Fock} imply high degeneracy of its energy levels.
An external perturbation breaks the symmetry and lifts the degeneracy~\cite{Landau}.
The nature of the perturbation can widely differ and can include, for example, external magnetic or 
electric fields~\cite{Gallagher,Sivalertporn,Popov,WintgenPRL,Tureja}, the influence of a third 
particle~\cite{Efimov,Faddeev,Drake}, or an artificial confining potential~\cite{Ekimov,Davies,Alferov}.
The latter is especially important for quantum optics, where confinement allows one to trap two-level 
and many-level systems into spatial lattices~\cite{Haroche,Berloff}.
Thus, quantum confinement is an essential tool which, on the one hand, makes it possible to localize 
quantum systems and, on the other hand, allows for the interaction of the localized 
objects~\cite{Schaefer}.

In semiconductor physics, quantum confinement is a basic way to specify the properties of radiative quantum systems~\cite{Kazarinov,Ekimov,Haug,Naka2018,Efros}. The strength of the confinement defines the distribution of energy levels and, thus, frequencies of the emitted light.
Artificially grown heterostructures of GaAs, GaN, CdTe, and other materials provide a convenient framework for an experimental study of the influence of quantum confinement, namely the restriction of motion along one spatial dimension~\cite{Alferov,Fedichkin,Ferreira}.
The potential profiles of the grown structures are usually modeled by simple quantum well (QW) 
potentials of rectangular~\cite{Davies}, triangular~\cite{Ahn,Baines}, parabolic~\cite{ClarkCPC,Liew} or 
P\"{o}sch-Teller-like shapes~\cite{Landau}.
Although the effect of such confinement on single-particle states is well-understood~\cite{Davydov}, 
the effect of confinement on few-body systems is more complicated~\cite{Kezerashvili}.
The main point is that confinement of a two-body system requires solving a three-body Schr\"{o}dinger 
equation~\cite{Faddeev,Kostrykin,Lin,Kez2024}, in which the confining potential plays the role of the third 
particle~\cite{Efremov}.

The confinement regime depends on the interplay between the size of the Rydberg state and the size of the QW.
This means that, although for a given QW width the ground state of the size of the Bohr radius can 
be in a weak confinement regime (model of wide QW approaching the bulk crystal), a highly-excited 
Rydberg exciton with a large principal quantum number can be strongly confined (model of 
narrow QW). The point is that if the mean size of the highly-excited state is large enough, 
the state becomes squeezed in the QW.
Even for the simplest model potentials, the crossover from weak to strong quantum confinement 
cannot be treated analytically. However, there are limiting, exactly solvable cases in the vicinity of 
which the perturbative treatment is accurate.
For example, in the case of weak confinement, the Coulomb attraction dominates: the energy difference between quantum-confined energy levels is smaller than that between Coulomb bound states. Here, the quantum confinement can be treated as a small perturbation.
The other limiting case is the strong confinement when, in turn, the Coulomb potential is treated as a small perturbation.
In this case, the gaps between quantum-confined energies are much larger than the Rydberg energy of the Coulomb interaction.
A crossover from strong to weak confinement and further to the bulk crystal allows one to follow the evolution of the energy levels, which provides an important insight for experimentalists to predict the energetic properties of their grown structures.

Practical examples of the effect of quantum confinement are cascade lasers. They include the well-known 
quantum cascade lasers~\cite{Kazarinov} as well as recently realized bosonic cascade 
lasers~\cite{Liew}. 
GaAs-based heterostructures with a parabolic-like potential profile can be grown by a gradual change of 
the alloy concentration~\cite{Tzimis,Trifonov}. As a result, a series of equidistant quantum-confinement energy levels appear, which allow the excitonic transitions to be amplified by the bosonic 
stimulation of radiative transitions between levels in a cascade. Such a parabolic structure is a key 
element of the proposed bosonic cascade laser designed to generate THz radiation~\cite{Liew}. This 
concept has thus far been realized in a variety of different structures~\cite{Ballabio,Deimert,Chen}.

Stronger confinement allows one to obtain more stable quantum systems, that can operate at higher 
temperatures. For example, confining an exciton to a two-dimensional plane increases its ground state binding energy 
by a factor of four~\cite{Ivchenko}. In this regard, cuprous oxide is a promising material due to large 
binding energies of quasiparticles already in the bulk. Indeed, bulk Cu$_{2}$O crystals have been shown
to possess many sharp exciton resonances including highly-excited Rydberg states~\cite{Kazimierczuk}.
The bulk exciton states, the impact of the upper subbands as well as external fields on Rydberg 
excitons in Cu$_{2}$O, and a variety of many-body effects have been studied in 
detail~\cite{Luttinger55,Luttinger56,SuzukiHensel74,Scheel1,Scheel2,Scheel3,Schweiner16b,Joerg2,Rommel2021,Joerg3,Thewes15,Heckoetter2,Walther1,Walther2}.
Moreover, artificially grown high-quality cuprous oxide crystals are currently being 
fabricated~\cite{Giessen,Lynch,NakaPRL,Sekkat,Trinkler,Alaeian,Giessen2022}.
The grown samples of predefined size open up the possibility to confine excitons in these structures 
similar to QWs. The produced samples already allowed to experimentally study the radiative decay rates 
of Rydberg excitons in thin Cu$_{2}$O films~\cite{Naka2018} as well as the crossover from the excitonic 
superradiance regime to the polaritonic long-range propagation~\cite{NakaPRL}.
In this context, the systematic studies of Rydberg exciton states, their size-dependent properties in cuprous oxide QW-like structures become ever more important.

In this article, we study the two-band (or hydrogen-like) model of Rydberg excitons in Cu$_{2}$O-based QWs in different regimes ranging from weak to strong confinement.
We analyze the dependence of bound-state energies on a single parameter characterizing the strength of the confinement.
We begin our investigation by discussing parabolic confinement along one spatial 
direction, for example along the growth axis. In this case, the Kohn theorem~\cite{Kohn,Peeters} allows 
us to significantly simplify the treatment of the problem and to follow the evolution of energy levels 
during the crossover from weak to strong confinement.
The case of weak confinement is studied numerically using an expansion of the wave function over the 
Coulomb-Sturmian basis. We show that in such a case the distribution of the energy levels is determined by the 
degree of orientation of the corresponding wave functions over the confining direction.
As the Coulomb-Sturmian basis is appropriate only for weak confinement, the evolution of the exciton energy levels during the crossover from weak to strong confinement has to be investigated by alternative numerical methods such as a finite-difference discretization~\cite{Khramtsov} or a more precise B-spline expansion~\cite{deBoor,Bachau} of the exciton wave 
function. One then observes Rydberg energy levels as well as their crossings and avoided crossings during 
the evolution as the confinement becomes gradually stronger.
Moreover, one observes that the energy levels, distributed over the value of the principal quantum 
number $N$ in case of the weak confinement, change their order to a distribution over the value of the 
magnetic quantum number $m$ for strong confinement.

After having understood the general picture of the evolution of the energy levels for parabolic 
confinement, we turn to study a rectangular QW structure.
To this end, we use a two-band model of the Rydberg exciton in QW~\cite{Ivchenko} with cuprous 
oxide material parameters~\cite{Schweiner16b}, thus simulating the Cu$_{2}$O thin film sandwiched between 
vacuum or air. We disregard features of the band structure,  
i.e., effects of the spin-orbit split band and the 
nonparabolicity~\cite{Ivchenko,KhrantsovBelov}.
The effects of the complex valence band structure in bulk cuprous oxide have been discussed 
in detail before~\cite{Schweiner16b,Joerg2,Luttinger55,Luttinger56,SuzukiHensel74} and are thus 
only briefly mentioned here. We would like to point out that the overall structure of the excitonic 
Rydberg series can be well explained within a hydrogen-like two-band model, and only the details of 
the spectra require the consideration of the complete valence band structure.
We use the pure Coulomb potential to observe the 
general structure of energy levels and further compare it with the results obtained with the 
Rytova-Keldysh potential~\cite{Rytova,Keldysh}, taking into account the dielectric contrast in the 
QW and in the barriers~\cite{Thoai}.
Our model leads to the Schr\"{o}dinger equation with a Hamiltonian that produces an energy spectrum with many quantum-confinement subbands and different branches of continua.
In contrast to a parabolic potential, for a rectangular confining potential one cannot separate 
variables, and one has to numerically solve the full three-dimensional Schr\"{o}dinger 
equation~\cite{Belov2019}.
In this regard, we expand the exciton wave function over a basis of B-splines~\cite{deBoor,Bachau}.
Due to the large number of quantum-confined energy levels, we restrict our attention to states 
characterized by magnetic quantum numbers $m=0,\pm 1$, and calculate ground and several excited energy 
levels below the electron-hole ($eh$) scattering threshold.
The behavior of energy levels for higher magnetic quantum numbers can then be qualitatively understood 
from the above-mentioned model of parabolic confinement.
We observe the evolution of the energy levels of Rydberg excitons during the crossover from a narrow to a wide QW. Furthermore, we show that in these limiting cases the energies can be easily calculated by 
expanding the wave function over quantum-confined states. The intermediate range of the QW widths, 
however, cannot be treated by these techniques, and one has to resort to numerical B-spline solutions.
We compare our computational results with the peak positions in the photoluminescence spectrum of Cu$_{2}$O QW-like structure presented in Ref.~\cite{Naka2018}.

In this article, we focus on the bound states of $eh$ pairs in QW. In fact, a confinement along one 
dimension produces many quantum-confined subbands. For strong confinement, the confinement energies are 
large. Below each value of the sum of electron and hole confinement energies, a proper Rydberg series 
of $eh$ energies appears.
In QW systems, the $eh$ bound states are located below the lowest scattering threshold, 
i.e., below the sum of the lowest quantum-confined energies of the disjoint electron and hole in the
QW~\cite{Belov2019}. The higher-lying Rydberg series lie in the Coulomb continuum of the lower subbands, 
and are thus resonance states. They are characterized by additional linewidth broadening due to their 
finite lifetime, as the electron can scatter off the hole in the QW plane even in case of infinite QW 
barriers~\cite{Belov2022}. As the resonant states require a yet more elaborate 
treatment~\cite{Moiseyev,Scheuler2024}, in this article we focus solely on the bound $eh$ states, characterized by 
their square-integrable wave function.

The article is organized as follows. We begin in Sec.~\ref{sec:1Dconfinement} by setting the scene 
for describing Rydberg excitons in structures that provide a spatial confinement in one dimension.
We then elaborate on two illustrative examples, a parabolic confinement (Sec.~\ref{sec:parabolic}) 
and a rectangular confinement (Sec.~\ref{sec:rectangular}). We specify the particular 
Cu$_{2}$O-based QW structure and describe numerical methods used to compute the exciton spectrum 
there. A detailed discussion of the obtained results is provided in Sec.~\ref{sec:resultsP} and 
Sec.~\ref{sec:resultsR} for parabolic and rectangular confinement, respectively.
Section~\ref{sec:RK} describes the results obtained with different dielectric constants in 
the QW and in the barriers.
A brief discussion of the effects of the crystal environment on Rydberg excitons in QWs is presented in Sec.~\ref{sec:valence}.
Numerical details are given in the Appendix.

\section{Rydberg excitons confined in one spatial dimension}
\label{sec:1Dconfinement}
In the two-band (or hydrogen-like) approximation~\cite{Ivchenko,Haug}, electron and hole 
dispersions are assumed to be parabolic, and the bound $eh$ states appear solely due to 
the Coulomb attraction.
However, the quantum confinement along one axis significantly complicates the well-known
Rydberg-like series~\cite{Scheel1,Schweiner16b} of energy levels.

The quantum states of the $eh$ pair confined in the $z$ direction are defined by 
the eigenstates of the Hamiltonian
\begin{equation}
H=\frac{\mathbf{p}_{e}^{2}}{2m_{e}}+\frac{\mathbf{p}_{h}^{2}}{2m_{h}}-\frac{e^{2}}{\epsilon|\mathbf{r}_{e}-\mathbf{r}_{h}|}+V_{e}(z_{e})+V_{h}(z_{h}),
\label{eq:Hamiltonian}
\end{equation}
where $m_{e}$ and $m_{h}$ are the effective masses of the electron and the hole in the 
semiconductor, respectively.
The variables $z_{e}$ and $z_{h}$ are the coordinates along the confinement direction, whereas 
$\rho$ is the distance between electron and hole in the QW plane.
The confining potentials $V_{e}(z_{e})$ and $V_{h}(z_{h})$ break the spherical symmetry of the 
problem, reducing it to mere translational invariance in the QW plane.
Note that we use a simple two-band model for the kinetic energy of the electron and hole in 
Eq.~\eqref{eq:Hamiltonian}, which means that the effects of the complex valence band structure on 
Rydberg excitons in QWs are neglected. Brief remarks on these effects are given in 
Sec.~\ref{sec:valence}.

In terms of the center-of-mass coordinate $\mathbf{R}=(m_{e} \mathbf{r}_{e}+m_{h} 
\mathbf{r}_{h})/(m_{e}+m_{h})$, and momentum $\mathbf{P}=\mathbf{p}_{e}+\mathbf{p}_{h}$, as well as 
the relative coordinate $\mathbf{r}=\mathbf{r}_{e}-\mathbf{r}_{h}$ and momentum $\mathbf{p}=(m_{h} 
\mathbf{p}_{e}-m_{e} \mathbf{p}_{h})/(m_{e}+m_{h})$, the $eh$ Hamiltonian reads
\begin{equation}
H=\frac{\mathbf{P}^{2}}{2M}+\frac{\mathbf{p}^{2}}{2\mu}-\frac{e^{2}}{\epsilon|\mathbf{r}|}+V_{e}(Z+\beta z)+V_{h}(Z-\alpha z).
\end{equation}
Here $M=m_{e}+m_{h}$ is the total exciton mass, $\mu=(m_{e}^{-1}+m_{h}^{-1})^{-1}$ is the reduced 
mass, and $\alpha=m_{e}/(m_{e}+m_{h})$ and $\beta=m_{h}/(m_{e}+m_{h})$ are normalized electron and hole 
masses, respectively.

Due to translational invariance in the QW plane, the center-of-mass motion in this plane can 
be separated. Furthermore, due to rotational symmetry the angular momentum component along the 
$z$-direction is conserved. This implies that the unknown wave function can be written as
$\psi(Z,z,\rho) \, e^{im\phi} / {\sqrt{2\pi}}$ where $m\in \mathbb{Z}$ is the magnetic quantum 
number and $\phi$ is the polar angle in the QW plane.
For instance, the value $m=0$ defines cylindrically symmetrical solutions.
As a result, 
the Hamiltonian of the nontrivial $eh$ motion is given by
\begin{align}
\nonumber &H(Z,z,\rho)=\frac{P^{2}_{Z}}{2M}+\frac{p^{2}_{z}}{2\mu}-\frac{\hbar^{2}}{2\mu}\left( \frac{\partial^{2}}{\partial \rho^{2}} + \frac{1}{\rho} \frac{\partial}{\partial \rho} -\frac{m^{2}}{\rho^{2}} \right) \\
&-\frac{e^{2}}{\epsilon \sqrt{\rho^{2}+z^{2}}}+V_{e}(Z+\beta z)+V_{h}(Z-\alpha z).
\label{ex3D}
\end{align}
Returning to absolute $z$-coordinates of the electron and the hole instead of the center-of-mass and 
relative ones $(Z,z)$, the last equation can be written as~\cite{Belov2019}
\begin{align}
\nonumber &H(z_{e},z_{h},\rho)=-\frac{\hbar^{2}}{2m_{e}} \frac{\partial^{2}}{\partial z_{e}^{2}}-\frac{\hbar^{2}}{2m_{h}} \frac{\partial^{2}}{\partial z_{h}^{2}}\\
\nonumber &-\frac{\hbar^{2}}{2\mu}\left( \frac{\partial^{2}}{\partial \rho^{2}} + \frac{1}{\rho} \frac{\partial}{\partial \rho} -\frac{m^{2}}{\rho^{2}} \right) \\
&-\frac{e^{2}}{\epsilon \sqrt{\rho^{2}+(z_{e}-z_{h})^{2}}}+V_{e}(z_{e})+V_{h}(z_{h}).
\label{eq:3Deq}
\end{align}
The energies $E$ of the electron-hole pairs are thus the solutions of the Schr\"{o}dinger equation
\begin{equation}
\label{SchEq}
H(z_{e},z_{h},\rho) \psi(z_{e},z_{h},\rho) = E \psi(z_{e},z_{h},\rho).
\end{equation}

The energies of the Rydberg exciton states can be defined with respect to the sum of the lowest quantum-
confined energies $E_{e1}+E_{h1}$ of electron and hole in the confinement potentials $V_{e}(z_{e})$ and 
$V_{h}(z_{h})$, respectively.
This sum specifies the lower boundary of the continuum, i.e., the lowest scattering threshold.
Below $E_{e1}+E_{h1}$ there are only bound states, and above this threshold the resonant states 
(of the particular parity) appear. Thus, the exciton binding energy is defined as
\begin{equation}
\label{eqEb}
E_{b}=E_{e1}+E_{h1}-E.
\end{equation}

\section{Parabolic confinement}
\label{sec:parabolic}
We begin with an illustrative example of parabolic confinement over one axis that is, to a certain 
extent, analytically tractable in different confinement regimes and which has been realized in various
semiconductor heterostructures~\cite{Bajoni07,Balili08,Amo10,Wertz10,Tosi12}.
We assume the confining potential to be of the 
form $V_{e,h}(z)=m_{e,h}\Omega^2 z^{2}/2$, i.e., it restricts the motion of the exciton in the $z$ 
direction. With this form of the confining potential, the center-of-mass and relative motions can still
be exactly separated~\cite{Kohn,Peeters}. That is, the wave function for the Hamiltonian~(\ref{ex3D}) can be written as
$\psi(Z,z,\rho)=\Psi(Z)\Phi(z,\rho)$. Here, $\Psi(Z)$ is the solution of the one-dimensional 
Schr\"odinger equation for the exciton as a whole, trapped in the harmonic potential~\cite{Landau}
\begin{equation}
\left[ -\frac{\hbar^{2}}{2M} \frac{d^{2}}{d Z^{2}}+\frac{M \Omega^{2}}{2} Z^{2} \right] \Psi(Z) = 
E_{Z} \Psi(Z).
\end{equation}
The function $\Phi(z,\rho)$ is, in turn, the solution of the two-dimensional equation for the relative 
$eh$ motion,
\begin{widetext}
\begin{equation}
    \left[ -\frac{\hbar^{2}}{2 \mu}
    \left( \frac{\partial^{2}}{\partial \rho^{2}} + \frac{1}{\rho} \frac{\partial}{\partial \rho} - 
    \frac{m^{2}}{\rho^{2}} \right)
    - \frac{e^{2}}{\epsilon \sqrt{\rho^2+z^2}}
    -\frac{\hbar^{2}}{2 \mu} \frac{\partial^{2}}{\partial z^{2}} + \frac{\mu \Omega^{2}}{2} z^{2} 
    \right] \Phi(z,\rho)= E_{z}\Phi(z,\rho).
\label{exParabolic}
\end{equation}
\end{widetext}

The $eh$ bound states are located below the lowest quantum confinement energy of the parabolic 
potential, that is, below $E_{\mu1}=\hbar\Omega/2$. Thus, the binding energies~\eqref{eqEb} are defined 
as
\begin{equation}
\label{eqEBparabolic}
E_{b}=E_{\mu1}-E_{z}.
\end{equation}
Hence, the separation of variables due to parabolic confinement reduces the dimensionality of 
the equation of motion for the relative coordinate when compared to Eq.~\eqref{SchEq}, which
significantly simplifies the theoretical investigation. Numerical solution of Eq.~\eqref{exParabolic} 
allows one to study the evolution of the energy spectrum during the crossover from the exciton 
in a bulk crystal (weak confinement, 3D exciton) to the exciton confined in a thin film (strong 
confinement, 2D exciton).

\subsection{Weak parabolic confinement}
In the case of weak parabolic confinement, the solution of Eq.~\eqref{exParabolic} can be obtained by 
taking the weak confinement as a perturbation. In this case, the unperturbed equation is a 
hydrogen-like problem, and thus the expansion of the wave function is conveniently performed over 
hydrogen-like basis functions. To be precise, the Schr\"{o}dinger equation for the hydrogen-like 
problem with parabolic confinement along the $z$-axis reads
\begin{equation}
\label{weakParabolic}
    \left[ -\frac{\hbar^2}{2\mu} \Delta - \frac{e^2}{\epsilon} \frac{1}{r} + \frac{\mu \Omega^{2}}{2} 
    z^{2} \right] \psi(\mathbf{r})= E_{z} \psi(\mathbf{r}).
\end{equation}
The parabolic potential can be expressed in terms of the radial variable and a linear combination of 
spherical harmonics~\cite{Varshalovich} $Y_{l}^{m}(\theta,\phi)$ as
$$
\frac{\mu \Omega^{2}}{2} (r \cos{\theta})^{2} = \frac{\mu \Omega^{2} r^2}{2}
\left( \sqrt{\frac{4 \pi}{9}} Y_{0}^{0} (\theta,\phi) + \sqrt{\frac{16 \pi}{45}}Y_{2}^{0} (\theta,\phi) 
\right).
$$
In spherical coordinates $(r,\theta,\varphi)$, the three-dimensional Coulomb-Sturmian basis is 
convenient for expansion of the wave function as the matrix elements of the kinetic and 
Coulomb potential terms are known~\cite{Wintgen,Clark}. With radial $n_{r}$, orbital $l$, and magnetic 
$m$ quantum numbers, the basis functions are naturally defined as
\begin{equation}
\label{CSbasis}
\Phi_{n_{r}lm}(r,\theta,\varphi;\lambda)=\frac{1}{r} \psi_{n_{r},l}(r;\lambda) Y^{m}_{l}(\theta,
\varphi),
\end{equation}
where the Coulomb-Sturmian functions read
\begin{eqnarray}
\label{CSfunctions}
    \psi_{n_{r},l}(r;\lambda) &=& \left(\frac{n_{r}!}{(2l+n_{r}+1)!} \right)^{1/2} \mathrm{e}^{-\lambda 
    r/2}
    \nonumber \\ && \times (\lambda r)^{l+1} L_{n_{r}}^{2l+1}(\lambda r).
\end{eqnarray}
Here, $L_{n_{r}}^{2l+1}(\lambda r)$ are the generalized Laguerre polynomials~\cite{AS} and $\lambda$ is 
a variational parameter. Alternatively, one can redefine the functions in terms of the principal 
quantum number via a substitution $N=n_{r}+l+1$.

In this basis, the Coulomb potential is diagonal, viz.
\begin{eqnarray*}
&& \int \Phi_{n'_{r},l'}^{m'}(r,\theta,\varphi;\lambda) \; \frac{1}{r} \;
\Phi_{n_{r},l}^{m}(r,\theta,\varphi;\lambda) d\mathbf{r} 
\\ && = \delta_{n_{r}n'_{r}} \delta_{ll'} \delta_{mm'}\,.
\end{eqnarray*}
The Laplace operator in the Coulomb-Sturmian basis is, in turn, tridiagonal.
Correcting the typos in Ref.~\cite{Wintgen}, the matrix elements of the Laplace operator are
\begin{eqnarray*}
&& \int \Phi_{n'_{r},l'}^{m'}(r,\theta,\varphi;\lambda) \; \Delta \; \Phi_{n_{r},l}^{m}(r,\theta,
\varphi;\lambda) \;
d\mathbf{r}   \\ &&=
(-1)^{n_{r}+n'_{r}+1} \; \frac{\lambda^2}{4} \int \Phi_{n'_{r},l'}^{m'}(r,\theta,\varphi;\lambda) \; 
\mathbf{I} \; \Phi_{n_{r},l}^{m}(r,\theta,\varphi;\lambda) d\mathbf{r}.
\end{eqnarray*}
The matrix elements of the identity operator $\mathbf{I}$ are
\begin{eqnarray*}
&& \int \Phi_{n'_{r}.l'}^{m'}(r,\theta,\varphi;\lambda) \; \mathbf{I} \; \Phi_{n_{r},l}^{m}(r,\theta,
\varphi;\lambda) \; d\mathbf{r} \\ && =
\delta_{ll'} \delta_{mm'} \frac{1}{\lambda}
\left\{
\begin{array}{ll}
    2(n_{r}+l+1), &\, n_{r}=n'_{r} \\
    -\sqrt{(n_{r}+1)(n_{r}+2l+2)}, & \, n_{r}=n'_{r}+1
  \end{array}
\right. .
\end{eqnarray*}

In order to calculate the matrix elements of the parabolic potential, the angular and radial 
dependencies are evaluated separately. The angular part gives exact expressions including the Wigner-3j 
symbols~\cite{Varshalovich},
\begin{eqnarray}
\label{EqML}
&&
\nonumber \langle Y_{l'}^{m'}|\cos^{2}{\theta}|Y_{l}^{m} \rangle =
\\ \nonumber &&  (-1)^{m'} \sqrt{\frac{(2l'+1)(2l+1)}{4\pi}}
\sqrt{\frac{4\pi}{9}}
  \begin{pmatrix}
    l' & 0 & l \\
    0 & 0 & 0
  \end{pmatrix}
  \begin{pmatrix}
    l' & 0 & l \\
    -m' & 0 & m
  \end{pmatrix}
 \\ && 
  +(-1)^{m'} \sqrt{\frac{5(2l'+1)(2l+1)}{4\pi}}
\sqrt{\frac{16\pi}{45}}
  \begin{pmatrix}
    l' & 2 & l \\
    0 & 0 & 0
  \end{pmatrix}
  \begin{pmatrix}
    l' & 2 & l \\
    -m' & 0 & m
  \end{pmatrix}.
\nonumber\\
\end{eqnarray}
The radial part can also be computed analytically as
\begin{eqnarray}
\nonumber && \int\limits_{0}^{\infty} \phi_{n'_{r},l'}(r,\lambda) r^{4} \phi_{n_{r},l}(r,\lambda) \; dr 
\\
\nonumber && =\frac{\lambda^{2}}{8}
\sqrt{\frac{n_{r}! n'_{r}!}{(n_{r}+2l+1)! (n'_{r}+2l'+1)!}} \\
\nonumber && \times
\sum_{i=0}^{n_{r}} \sum_{j=0}^{n'_{r}} (-1)^{i+j}
  \begin{pmatrix}
    n_{r}+2l+1 \\
    n_{r}-i
  \end{pmatrix}
  \begin{pmatrix}
    n'_{r}+2l'+1 \\
    n'_{r}-j
  \end{pmatrix}\\
&& \times
  \frac{(i+j+l+l'+4)!}{i! j!}.
\label{eqSum}
\end{eqnarray}
Moreover, the sums on the rhs of Eq.~(\ref{eqSum}) can be evaluated exactly as shown in the Appendix of 
Ref.~\cite{Clark}.

In practical calculations, when the parabolic confinement is weak (small $\Omega$), one has to 
take into account basis functions with only a few quantum numbers to obtain accurate energies. For 
stronger confinement (larger $\Omega$), many more basis functions are needed to achieve the same 
precision. As a result, despite the sparse block structure of the Hamiltonian in the Coulomb-Sturmian
basis, the number of blocks is given by the maximal value of the included quantum numbers and increases 
as $O(l_{\max}^{2}N_{\max})$. This limits the applicability of the proposed scheme and makes the 
solution unpractical for the strength of the confinement
$A=\mu\Omega^{2}/2\gtrsim 0.1$~Ry/$a_{B}^{2}$.
%
\begin{figure}[tp!h]
\begin{center}
\includegraphics[angle=0, width=\columnwidth]{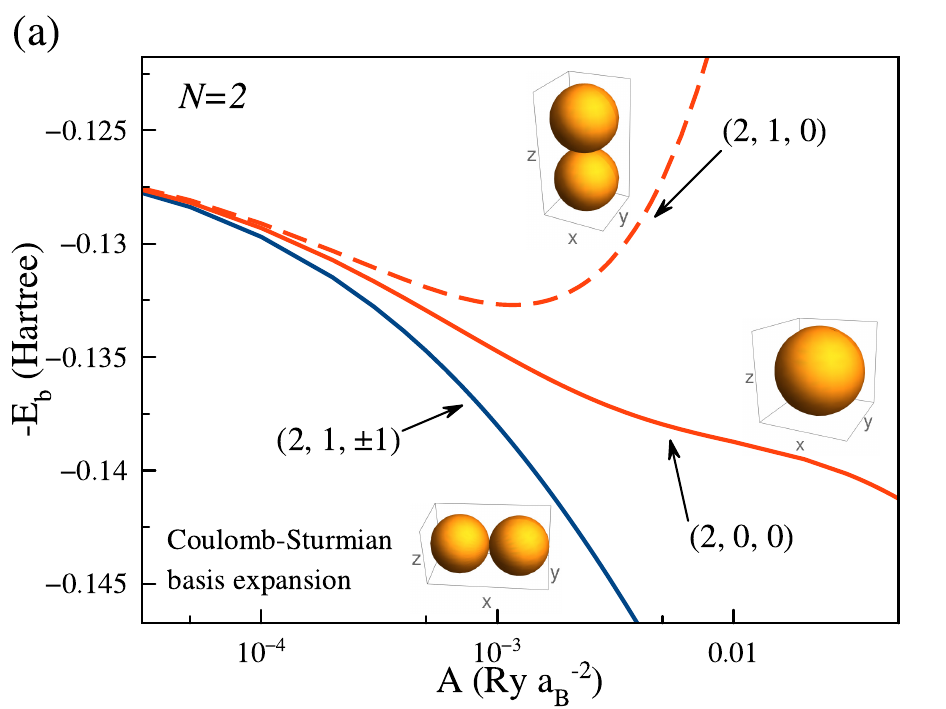}
\includegraphics[angle=0, width=\columnwidth]{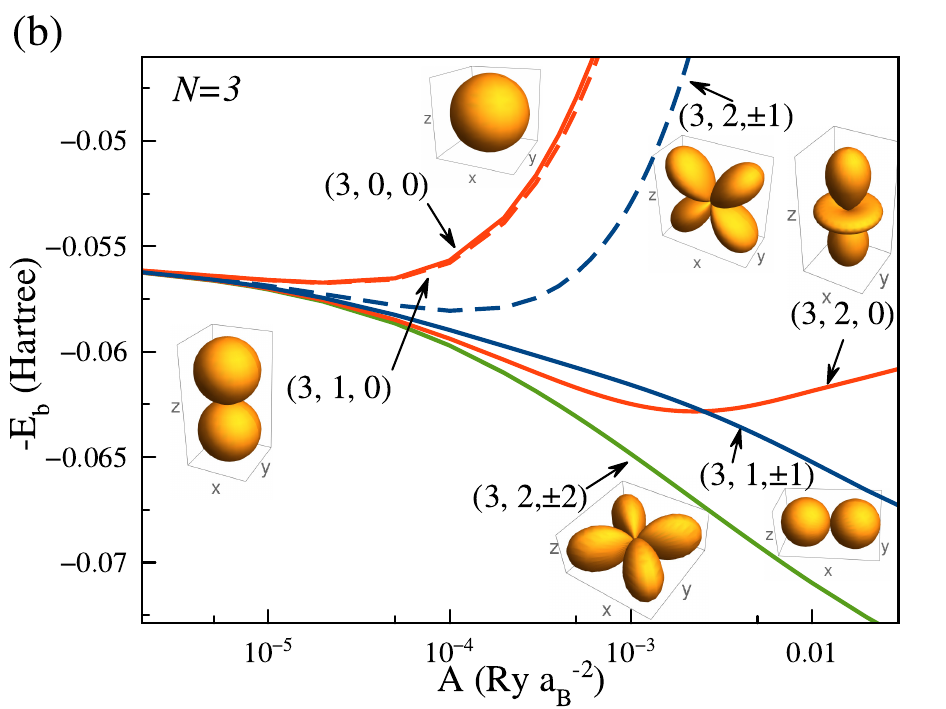}
\caption{Splitting of the energy levels as function of the strength of the parabolic confinement 
$A=\mu\Omega^{2}/2$~Ry/$a_{B}^{2}$ using Coulomb-Sturmian basis expansion.
(a) energies corresponding to principal quantum number $N=2$,
(b) $N=3$. The energy levels are characterized by three quantum numbers $(N,l,m)$,
where $N$ is the principal quantum number, $l$ is the orbital one, and $m$ is the magnetic one.
The energy levels with same $|m| \ge 1$ are degenerate and correspond to two different states.}
\label{figN2}
\end{center}
\end{figure}

\subsection{Strong confinement}
\label{SectStrongParabolic}
In general, Eq.~\eqref{weakParabolic} determines the energies for arbitrary strength
$A=\mu\Omega^{2}/2$~Ry/$a_{B}^{2}$
of the parabolic confinement. However, in case of the strong confinement (large 
$\Omega$), the procedure described in the previous section is impractical. 
In such a case, a straightforward numerical solution of Eq.~\eqref{exParabolic} is more convenient.
For a wide range of $\Omega$, it can be solved directly by the finite-difference 
method~\cite{Belov2022} or by an expansion of the wave function over a basis of Hermite 
functions~\cite{Hermite1,Hermite2} or B-splines~\cite{deBoor}.
Moreover, particularly for large $\Omega$, one can use the adiabatic approximation,
taking into account the stronger confinement along the $z$-direction than in the QW plane (along 
$\rho$). This approximation requires, first, a determination of the one-dimensional wave functions 
$\phi_{i}(z)$ and the corresponding energy levels of the parabolically confined exciton along the 
$z$-direction and, second, a solution of the coupled system of equations with the effective potential 
defined by the so obtained wave functions $\phi_{i}(z)$,
\begin{equation}
V^{\mathrm{eff}}_{ij}(\rho)=-\frac{e^{2}}{\epsilon}\int\limits_{-\infty}^{\infty} \frac{\phi_{i}(z) 
\phi_{j}(z)}{\sqrt{\rho^{2}+z^{2}}} \; dz.
\label{Veff}
\end{equation}

The adiabatic approach is accurate enough for large values of $\Omega$, when the nondiagonal terms of 
the effective potential~\eqref{Veff} are relatively small and, thus, can be treated as perturbations.
This means that the motion of the exciton is, to a good approximation, two-dimensional and the energies 
are close to those of the 2D Coulomb series.

\begin{figure}[tp!h]
\begin{center}
\includegraphics[angle=0, width=\columnwidth]{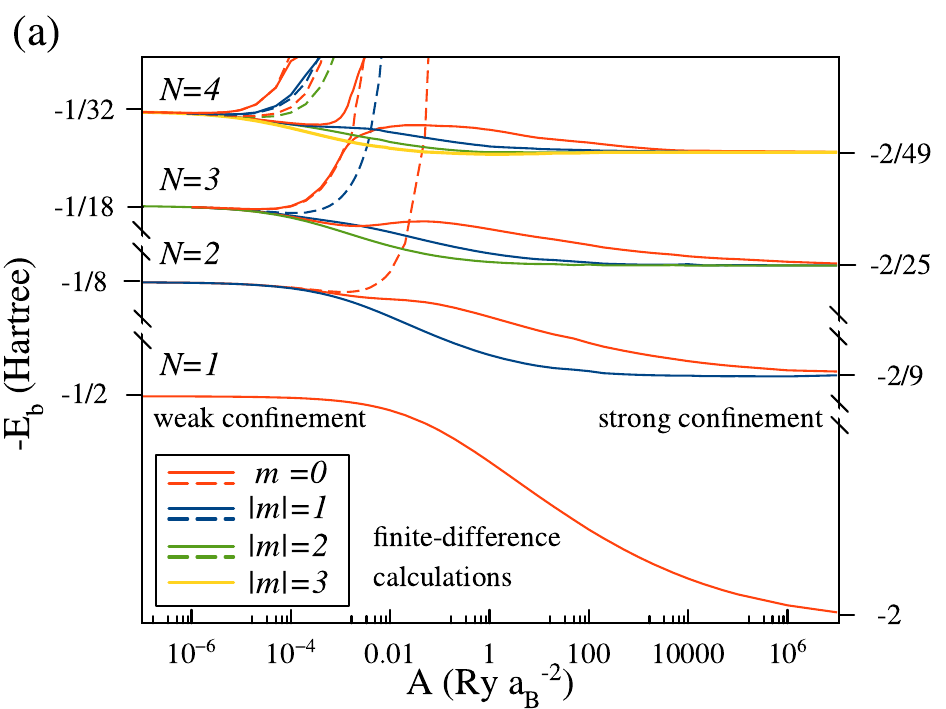}
\includegraphics[angle=0, width=\columnwidth]{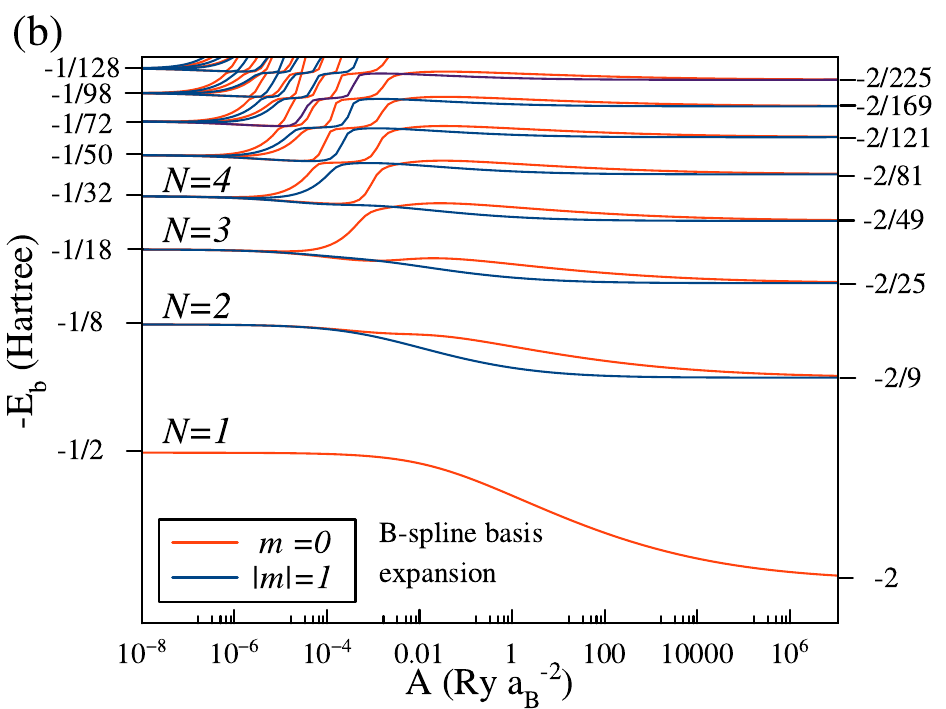}
\caption{The Rydberg series of exciton states depicted at the crossover from the weak to strong 
parabolic confinement along $z$. The parameter $A=\mu\Omega^{2}/2$~Ry/$a_{B}^{2}$ defines the strength of the confinement.
(a) Finite-difference calculations of several of the lowest energy levels. The solid 
lines correspond to the bound states which remain bound for both weak and strong confinement. The 
dashed lines show the states whose energies diverge during the crossover to the strong confinement. (b) 
Results of B-spline expansion computations of the excited states. The energy levels with $|m|\le 1$ are 
only shown for better visibility.}
\label{parabolic3Dto2D}
\end{center}
\end{figure}

\subsection{Results for parabolic confinement}
\label{sec:resultsP}

In this section, we present the calculation of the Rydberg series of energy levels of excitons in
parabolic quantum wells for different strengths of the confinement. The effective masses for the 
electron and hole in cuprous oxide, $m_{e}=0.99\,m_{0}$ and $m_{h}=0.69\,m_{0}$, and a dielectric 
constant $\epsilon=7.5$ have been used.
The variational parameter $\lambda$ in Eq.~\eqref{CSfunctions} is taken to be 1.
In our calculations, we studied only the $eh$ bound states. Their energy levels are below the 
scattering threshold $E_{\mu1}=\hbar\Omega/2$ of the lowest quantum confinement energy of a particle of 
the mass $\mu$, see Eq.~\eqref{exParabolic}, in a parabolic QW along the $z$ axis.
As a result, it is convenient to characterize the calculated energy levels by their binding 
energies~\eqref{eqEBparabolic}.
Depending on the strength $A=\mu\Omega^{2}/2$~Ry/$a_{B}^{2}$ of the parabolic confinement, different 
methods to evaluate the energies of the exciton states have been employed. 

For weak confinement, that is, for small $A$ (and thus small $\Omega$), the Coulomb-Sturmian 
basis~\eqref{CSfunctions}, together with the expansion over spherical harmonics allowed us to 
precisely calculate the dependence of the Rydberg energies on the confinement strength as an external 
perturbation. In the absence of the perturbation, the Coulomb energy levels are highly degenerate. If 
a weak perturbation is introduced, this degeneracy is lifted.
In Fig.~\ref{figN2}, we show the gradual increase in the splitting of the energy levels with an 
increase in the strength, $A$, of the weak perturbation.
Their shifts to lower or higher energies are determined mainly by the order of the alignment of the 
wave functions over the confining direction.

The quantum states can be characterized by three quantum numbers $(N,l,m)$, where $N$ is the principal 
quantum number, $l$ is the orbital quantum number, and $m$ is the magnetic one. For example, in 
Fig.~\ref{figN2}, the states corresponding to principal quantum numbers $N=2$ and $3$ are shown as 
function of the confinement strength $A$.
For $N=2$, taking into account the degeneracy of the states $(2,1,\pm 1)$, there are just three energy 
levels for four bound states, see Fig.~\ref{figN2}~(a). Thus, the interpretation of the results is 
rather straightforward~\cite{Glazov2018}. One observes that the energy of the state $(2,1,0)$ 
dramatically increases with growing $A$. The reason for this behavior is that the wave function of 
this state is essentially aligned along the $z$-axis. The squeezing of the wave function induces the 
rapid energy increase. The wave functions of the other three states are not aligned along $z$, so their 
energies do not change drastically. This is also seen in Fig.~\ref{parabolic3Dto2D}~(a), where the 
dependence of several of the lowest energy levels on a wide range of strengths $A$ is shown. 
Moreover, one observes that the wave function of the state $(2,0,0)$ is spherically symmetric, whereas 
the wave functions of the states $(2,1,\pm 1)$ are more aligned in the QW ($xy$) plane. As a result, the 
energy of the spherically symmetric state $(2,0,0)$ is above those of the states $(2,1,\pm 1)$.

For a principal quantum number $N=3$, see Fig.~\ref{figN2}~(b), the number of bound states is increased 
up to nine, so the interpretation of the results becomes somewhat more involved.
Moreover, in this case the distribution of states is additionally complicated by the coupling of states 
with the same $m$ and with the values of $l$ that differ by 2 [$m-m'=0$ and $l-l'=\pm 2$ in 
Eq.~\eqref{EqML}]. Thus, the corresponding energy levels exhibit the avoided crossings.
If this coupling were absent, then again the distribution of energy levels would be mainly defined by 
the measure of alignment of the wave function along the $z$-axis. The more the wave function is aligned 
along the confining direction, the higher the energy (lower binding energy) of the state. Therefore, 
the state $(3,1,0)$ would have the highest energy (lowest $|E_{b}|$) due to the largest $z$-component.
All other states would be located below that in energy until the lowest state $(3,2,\pm 2)$ whose
wave function lies predominantly in the QW plane.
However, due to their coupling, the energy levels of states with the same $m$ and $l-l'=\pm 2$ show
avoided crossings.
For $N=3$, there is a coupling of the angular dependencies of the states $(3,0,0)$ and $(3,2,0)$. This 
results in a state repulsion, making the state $(3,0,0)$ to be energetically highest. Furthermore, the 
states $(3,2,0)$ and $(3,1,\pm 1)$ cross at $A\approx 0.0025$~Ry$/a_{B}^{2}$.

The energies of the states $(2,1,\pm 1)$, $(3,1,\pm 1)$ and $(3,2,\pm 2)$ decrease with growing
strength $A$ due to their dominant alignment along the QW plane. In fact, the energies of these 
states transit from weak to strong confinement. For very weak confinement (3D exciton, 3D Coulomb 
potential), the states are degenerate and their binding energies are $-1/2N^{2}$~Hartree, whereas for 
the limiting case of strong confinement (2D exciton, 2D Coulomb potential), their energies are
$-1/2(N-1/2)^{2}$~Hartree.
Hence, the bound states aligned along the QW plane generally become more strongly bound as the strength 
$A$ increases~\cite{Ivchenko}. This is depicted in Fig.~\ref{parabolic3Dto2D} where the energy levels 
as functions of the strength of the confinement are shown.
One can see, for example, that the energy of the state $(2,1,\pm 1)$ decreases from $-1/8$~Hartree to 
$-2/9$~Hartree. The multiplicity of the 3D Coulomb energy level is $N^{2}$,
whereas for the 2D Coulomb potential it is $2N-1$.
Therefore, for $N=2$ we see that one of four states drastically increases its energy as the confinement 
growths and other three fill in the three empty vacancies in case of the strong confinement.

The Coulomb-Sturmian basis is not appropriate for a precise determination of the energies across the 
entire range of confinement strengths, covering both weak and strong regimes. For $N_{\max}\sim 100$ 
and $l_{\max}\sim 20$, the energy levels obtained using the Coulomb-Sturmian basis start to diverge at 
around $A \sim 0.1$~Ry/$a_{B}^{2}$. Therefore, for larger $A$ we employed the finite-difference 
method~\cite{Khramtsov,Belov2019} to accurately calculate the lowest energy levels, see 
Fig.~\ref{parabolic3Dto2D}~(a). Details are given in the Appendix~A. The finite-difference 
approximation allowed us to precisely calculate the energy levels up to $A\sim10^{6}$~Ry/$a_{B}^{2}$. For even 
stronger confinement, these results can also be obtained using the adiabatic approach, see 
Sect.~\ref{SectStrongParabolic}, because the confinement along $z$ dominates and the system is, to a large extent, two-dimensional.

Solving Eq.~\eqref{exParabolic} by the finite-difference method allows one both to determine several of the lowest energy levels for a large range of confinement strengths, $A=10^{-6}-10^{6}$~Ry/$a_{B}^{2}$, as well as their ordering.
Although the distribution for the excited Rydberg states is rather complicated, we can draw some important conclusions from the calculated spectra.
States with the same quantum numbers, that differ only by the sign of $m$, are degenerate.
For a given principal quantum number $N$, the two lowest degenerate energy levels have $|m|=l=N-1$. 
They remain there irrespective of the strength of the confinement, as they are mainly 
aligned along the QW plane and, moreover, have a minimal extent along the $z$ axis.
For weak confinement, one of the states with largest energy in a bundle is the state with $l=1$ and $m=0$. As it is mainly aligned along $z$, its energy rapidly diverges as $A$ increases.
Furthermore, for arbitrary $N$, the energies of states with $|m|=l-1$ diverge.
The spherical harmonics of these states are proportional to $z/r$ and thus the corresponding matrix element of the parabolic potential is positive and grows drastically with increasing $A$.

Generally, the quantum confinement leads to an increase of the binding energy~\cite{Ivchenko}.
However, one can see that instead some states increase their energies (diminish $E_{b}$) by crossing 
over higher-lying states with growing confinement, see Fig.~\ref{parabolic3Dto2D} (b).
During the crossover from weak to strong confinement, 
the energy levels for fixed $N$ are stacked above the lowest states $\left(N,N-1,\pm (N-1)\right)$ with 
decreasing magnetic quantum number $m$. This can be seen in Fig.~\ref{parabolic3Dto2D} (a) where for 
$N=4$, the energy levels of the bound states are ordered from bottom to top of the bundle as 
$|m|=3,2,1,0$.

Although the finite-difference method allows for a precise determination of lower-lying energy levels, 
it is less appropriate for the calculation of high-lying Rydberg exciton energy levels. The wave 
functions of the Rydberg states are significantly spatially extended and require using fine grids over 
a broad calculation domain. The B-spline basis expansion~\cite{Bachau} of the wave function in 
Eq.~\eqref{exParabolic} makes it possible to overcome these technical issues and to accurately estimate 
many excited energy levels for a wide range of confinement strengths.
We used B-splines of fifth order to represent the wave function, that led to the generalized eigenvalue 
problem which was solved by the QR algorithm~\cite{LAPACK}.
The details on the B-spline calculations are given in the Appendix~B.

In Fig.~\ref{parabolic3Dto2D}~(b), we show the results of the B-spline calculations of the energy 
levels for magnetic quantum numbers $|m|=0,1$.
One observes a crossover from weak to strong confinement as well as crossings and avoided crossings 
of different excited energy levels. As already mentioned, for weak confinement the energy levels are 
ordered according to the energies of the 3D Coulomb potential, $-1/2N^{2}$, whereas for strong 
confinement they are governed by the 2D Coulomb potential $-1/2(N-1/2)^{2}$.
In the intermediate region, for a given $N$ the splitting of energy levels into bundles with different 
values of $|m|$ occurs.
\begin{figure}
	\includegraphics[width=0.5\textwidth]{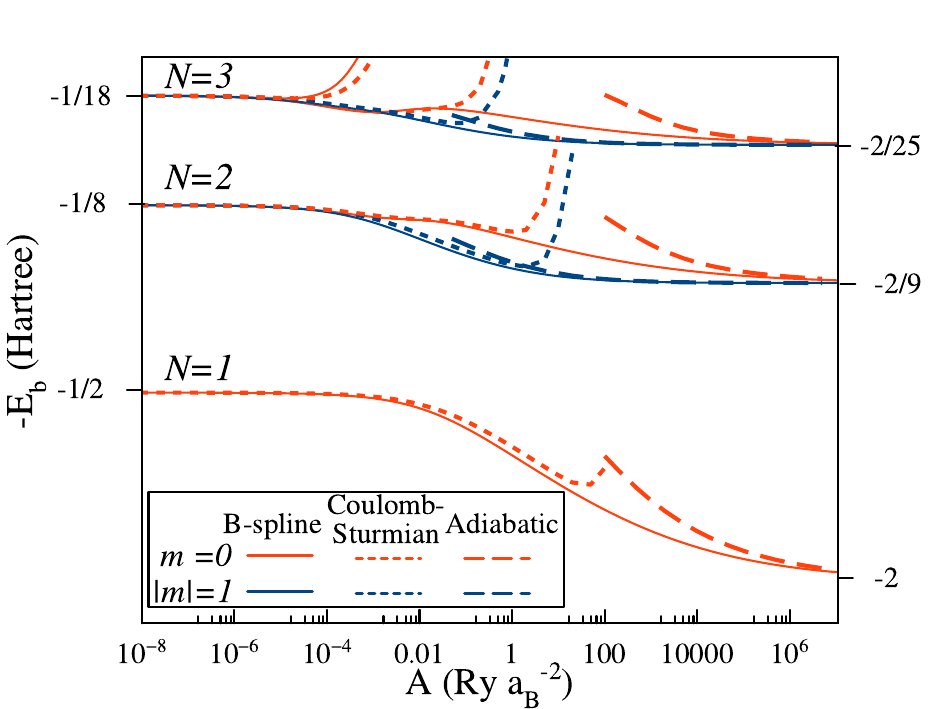}
	\caption{Exciton energies calculated by the B-spline expansion method (solid curves) in comparison with the ones obtained using the Coulomb-Sturmian (fine-dashed curves) and the adiabatic expansions (coarse-dashed curve).}
	\label{fig:parabolicUncertainty}
\end{figure}

The comparison of the exciton energies obtained by the B-spline expansion method and by the limiting 
expansions in weak and strong confinement regimes is shown in Fig.~\ref{fig:parabolicUncertainty}. The 
accurate B-spline solution is represented by the solid curves, whereas the energies calculated using 
the Coulomb-Sturmian and the adiabatic expansion methods are shown by dashed curves.
One can see that, although the limiting expansions are appropriate in the domain close to the 
corresponding limits, they are inaccurate for the whole studied range of confinement strengths.
Interestingly, deviations produced by the limiting models for energy levels with $m>0$ are noticeably 
smaller than for ones with $m=0$. They originate from different properties of the solutions of 
Eq.~\eqref{exParabolic} as $\rho\to 0$. Along the $\rho$ direction, the eigenfunctions of 
Eq.~\eqref{exParabolic} with $m>0$ vanish as $\rho\to 0$, whereas the solution for $m=0$ converges to a 
maximal constant value as $\rho\to 0$. The latter leads to a significant contribution to the potential 
term in Eq.~\eqref{Veff} in the interaction domain near $\rho=0$.
For $m>0$, as the solutions vanish as $\rho\to 0$, such contributions are much smaller and the 
deviations produced by the adiabatic and Coulomb-Sturmian expansions are reduced.

 \section{Rectangular confinement}
 \label{sec:rectangular}
In this section, we consider a rectangular confining potential
\begin{equation}
\label{recV}
V_{e,h}(z_{e,h}) = \left\{
  \begin{array}{lr}
    0 & \mbox{ if  } |z_{e,h}|<L/2 \\
    \infty & \mbox{ if  } |z_{e,h}| \ge L/2
  \end{array}
\right. ,
\end{equation}
where $L$ is the QW width.
In the case of a rectangular QW confinement, the center-of-mass and relative coordinates along the 
confinement direction ($z$ axis) in Eq.~\eqref{SchEq} can no longer be separated.
Therefore, in contrast to the previous section where the two-dimensional Schr\"odinger equation has 
been studied, here the complete three-dimensional Eq.~\eqref{ex3D} needs to be solved.
Nonetheless, we can adopt some of the approaches from the previous section.
We will treat the weak confinement as perturbation, with the three-dimensional Coulomb wave functions 
being the eigenfunctions of the unperturbed problem. Vice versa, the 3D Coulomb interaction will be 
considered as perturbation in the case of strong confinement.
For both approaches we solve the problem by expanding the three-dimensional wave function
into the respective tailored basis functions.
As a benchmark for comparison of the weak and strong confinement approaches, we will compare the 
results with those using the B-spline basis expansion. The B-spline basis is appropriate for computing 
the Rydberg energy levels for the whole range of studied QW widths, covering both weak and strong 
confinement regimes as well as the intermediate region, where the perturbative approaches are 
inaccurate.

\subsection{Weak confinement}
It has been suggested~\cite{Ivchenko} to treat the weak confinement of excitons by confining the 
motion of the exciton as a whole in terms of the center-of-mass coordinate $Z$, while the relative 
motion of electron and hole is not influenced by the confinement. Then, the nontrivial radial part of 
the wave function can be represented as an infinite sum~\cite{Belov2019}
\begin{equation}
\psi(Z,z,\rho) = \sum_{k,Nl}	
	c_{kNl}\;\phi_k(Z)\;\psi_{Nl}(\sqrt{\rho^{2}+z^{2}};\lambda),
	\label{eq:3DExpansion}
\end{equation}
where
\begin{equation*}
\phi_{k}(Z)= \sqrt{\frac{2}{L}} \left\{
  \begin{array}{lr}
    \cos(\pi k Z/L) & \text{ if } k=1,3,5,\ldots\\
    \sin(\pi k Z/L) & \text{ if } k=2,4,6,\ldots
  \end{array}
\right.
\end{equation*}
are the confinement states of a QW with infinite barriers. The relative motion of electron and hole is governed by the 3D Coulomb potential. As a result, $\psi_{Nl}(r;\lambda)$ are the Coulomb-Sturmian functions~\eqref{CSfunctions}, which have been introduced for weak parabolic confinement. 

We use the terms of the expansion~\eqref{eq:3DExpansion} to calculate the matrix elements of the exciton Hamiltonian $H(Z,z,\rho)$, see Eq.~\eqref{ex3D}, i.e., including both the confinement of center-of-mass and the relative motion:
\begin{equation}
\int\psi_{N'l'}(\rho,z;\lambda)\,\phi_{k'}(Z)\,
H(Z,z,\rho)
\,\psi_{Nl}(\rho,z;\lambda)\,\phi_{k}(Z)\,dZ\,dz\,d\rho
\label{eq:HExpansion}
\end{equation}
The zero boundary conditions are defined by the barriers of the confinement potentials 
$V_{e,h}(z_{e,h})$. As the confinement potentials depend both on $z$ and $Z$, the boundary conditions cannot 
be fulfilled by the basis wave functions (\ref{eq:3DExpansion}), which are separable in $z$ and $Z$. In 
analogy to the strong parabolic confinement, we make use of the adiabatic approach to obtain an 
effective $Z$-dependent potential by integrating over the relative coordinates
$(\rho,z)$.
Hereby, for each value of $Z$ the domain of integration over the $z$ coordinate is defined by the 
confinement potentials. As a result, the full integration 
domain in the $(Z,z)$ plane is reduced to a rhombic-like region~\cite{Belov2019}. 
Because of such an integration domain, one can no longer employ the recurrence relations of the 
Coulomb-Sturmian functions, but instead one has to numerically integrate them over $z$ and $\rho$.

\subsection{Strong confinement}
For strong confinement, one can assume a separate quantization of the electron and the hole motion 
along the confinement axis, leading to the confinement states $\phi_{i,j}(z_{e,h})$~\cite{Ivchenko}.
Moreover, the in-plane properties of the wave function are, in turn, mainly determined by the 2D 
Coulomb potential. As a result, the exciton wave function can be expanded as
\begin{equation}
\psi(\rho,\varphi,z_{e},z_{h};\lambda)
=\sum_{ijNm} c_{ijNm}\; \phi_{i}(z_e)\,\phi_{j}(z_h)\,\Phi_{Nm}(\rho,\varphi;\lambda).
\label{eq:2DExpansion}
\end{equation}

We introduce the 2D Coulomb-Sturmian functions as~\cite{Duclos}
\begin{equation}
    \Phi_{Nm}(\rho,\varphi;\lambda) = \frac{1}{\sqrt{\rho}}\, \phi_{Nm}(\rho;\lambda)\, 
    \frac{\mathrm{e}^{im\varphi}}{\sqrt{2\pi}},
\end{equation}
where $N$ and $m$ are the principal and magnetic quantum numbers, respectively, and
\begin{align*}
&\phi_{Nm}(\rho;\lambda)=\sqrt{\frac{(N-|m|-1)!}{(N+|m|-1)!}}
 \\& \times \mathrm{e}^{-\rho/\lambda}\left(\frac{2\rho}{\lambda}\right)^{|m|+1/2}L_{N-|m|-1}^{2|m|}
 \left(\frac{2\rho}{\lambda}\right).
\end{align*} 
Similar to the case of weak parabolic confinement, we expand the exciton wave function in the 
basis~\eqref{eq:2DExpansion} and find the tridiagonal matrix of the (in-plane) Laplace operator as
\begin{align*}
&\iint\Phi_{N'm'}(\boldsymbol{\rho};\lambda)\bigtriangleup_{\boldsymbol{\rho}}
\Phi_{Nm}(\boldsymbol{\rho};\lambda)\mathrm{d}\boldsymbol{\rho}\\
&=\frac{\delta_{m'm}}{\lambda}[-(N-\frac{1}{2})\delta_{N'N}-\frac{1}{2}(\sqrt{(N-m)(N+m)}
\delta_{N'(N+1)}\\&+\sqrt{(N+m-1)(N-m-1)}\delta_{N'(N-1)})],
\end{align*}
as well as of the identity operator
\begin{align*}
	\iint\Phi_{N'm'}(\boldsymbol{\rho};\lambda)&\Phi_{Nm}(\boldsymbol{\rho};\lambda)\mathrm{d}
 \boldsymbol{\rho}=\delta_{m'm}\,\lambda[(N-\frac{1}{2})\delta_{N'N}\\&-\frac{1}{2}(\sqrt{(N-m)(N+m)}
 \delta_{N'(N+1)}\\&+\sqrt{(N+m-1)(N-m-1)}\delta_{N'(N-1)})].
\end{align*}
The basis is diagonal with respect to the 2D Coulomb potential
\begin{equation}
	\iint\Phi_{N'm'}(\boldsymbol{\rho};\lambda)\frac{1}{\rho}\Phi_{Nm}(\boldsymbol{\rho};\lambda)
 \mathrm{d}\boldsymbol{\rho}=\delta_{N'N}\,\delta_{m'm}.
\end{equation}
This allows us to treat the difference to the 3D Coulomb potential as a perturbation
\begin{eqnarray}
&&\iint\Phi_{N'm'}(\boldsymbol{\rho};\lambda)\left(\frac{1}{\sqrt{\rho^2+(z_{e}-z_{h})^2}}
-\frac{1}{\rho}\right)\Phi_{Nm}(\boldsymbol{\rho};\lambda)\mathrm{d}\boldsymbol{\rho}
\nonumber \\
&&=V_{\mathrm{perturb}}^{N'N}(z_{e},z_{h}).
\label{VNNperturb}
\end{eqnarray}
Further integration of the QW confinement functions $\phi_{i,j}(z_{e,h})$ over $z_{e}$ and $z_{h}$ is 
straightforward, however, for the perturbation potential~\eqref{VNNperturb} this can only be done 
numerically. 

\subsection{Results for rectangular confinement}
\label{sec:resultsR}
Using the material parameters for the electron and the hole in cuprous oxide from 
Sect.~\ref{sec:resultsP}, the Rydberg energy and the Bohr radius of the lowest hydrogen-like exciton 
state in a bulk crystal are
$$
\text{Ry} = \frac{\mu e^4}{2 \epsilon^2 \hbar^2} = 98.35 \;\text{meV}, \quad a_{B}=\frac{\epsilon 
\hbar^2}{\mu e^2} = 0.976 \;\text{nm}.
$$
The latter value means that the Bohr radius of the exciton ground state is around 1~nm. Thus, for this 
state QW thicknesses $L$ (or sizes of the crystal) of the order of 10~nm and wider can be considered as 
a model of wide QWs. Such a model assumes that the Coulomb potential dominates, whereas quantization 
due to QW barriers can be treated as a small perturbation. 
As a result, for such QW widths the representation~\eqref{eq:3DExpansion} is justified.

In our study, we have computed the binding energies of the exciton states in the weak and strong 
confinement regimes using the expansions~\eqref{eq:3DExpansion} and~\eqref{eq:2DExpansion}, 
respectively. 
\begin{figure}
	\includegraphics[width=0.5\textwidth]{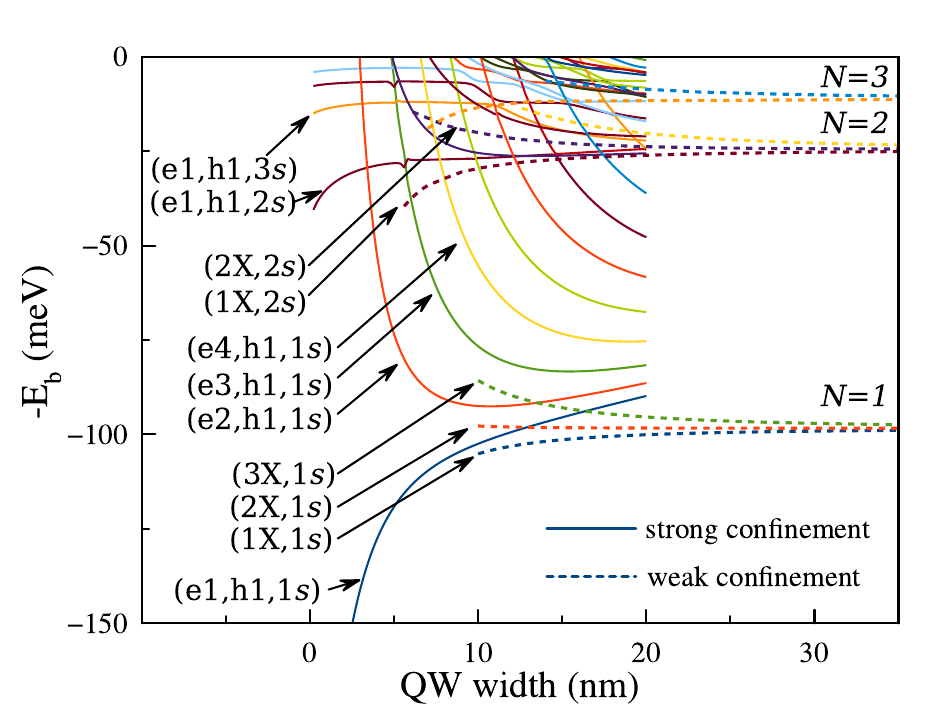}
	\caption{Exciton energy levels as a function of the QW width for $m=0$. The solid curves show the 
 energy levels obtained using the expansion over 2D Coulomb-Sturmian basis (strong-confinement 
 expansion), for the dashed ones we use the 3D basis (weak-confinement expansion). For clarity, higher 
 confinement states $(kX,Ns)$ are only partly shown. Selected classified energy levels are denoted.}
	\label{fig:QWExpansion}
\end{figure}
The calculated binding energies are presented in Fig.~\ref{fig:QWExpansion} in the respective limits of 
wide and narrow quantum wells. In these limits, both numerical methods show good convergence: the 
calculated Rydberg energy series correspond to the models of 3D Coulomb and 2D Coulomb confinement.
However, the convergence becomes worse if one moves away from the limiting regimes.
For example, if one considers the model of a narrow QW (strong-confinement expansion), and starts to 
gradually increase $L$, then the calculated results start to gradually diverge.
One can see in the figure that the energies of the strong-confinement expansion do not asymptotically 
converge to the Rydberg energies as $L\to\infty$. Instead of approaching the constant value, the ground 
and excited state energies diverge.
By contrast, the weak-confinement basis expansion~(\ref{eq:3DExpansion}) allows one to precisely 
calculate the exciton energies over a broad range of QW widths, when $L \gg a_{B}$. The calculated 
energies are shown in Fig.~\ref{fig:QWExpansion} by dashed curves. One can see the precisely obtained 
several lowest energies of the Rydberg series $N=1$, $2$, and $3$. For smaller $L$, this expansion 
gives inaccurate results.

The strong-confinement expansion~\eqref{eq:2DExpansion} makes it possible to directly assign
quantum numbers to the calculated energy levels. In the simplest case, our calculations are restricted 
to cylindrically symmetric solutions, $m=0$. Therefore, we can use three quantum numbers $(ei,hj,Ns)$ 
that refer to the $i$-th electron quantum-confinement state, the $j$-th hole quantum-confinement state 
and the 2D Coulomb principal quantum number $N$. 
Similarly, for the weak-confinement expansion~\eqref{eq:3DExpansion} it is convenient to introduce two 
quantum numbers: $k$ is for the $k$-th quantized state of the exciton as a whole over the $z$ axis, and 
$N$ is for the 3D Coulomb principal quantum number. They are collected in the doublet $(kX,Ns)$, where 
$X$ stands for the exciton. It is worth noting that, as resulting from Eq.~\eqref{eq:3DExpansion}, in 
general different orbital quantum numbers are allowed, although here we restrict ourselves to the 
simplest case of $l=0$. The assignments are exact in the corresponding limiting cases of strong and 
weak confinement.
However, when leaving the respective limits, they become approximate due to a mixing of the basis states.
\begin{figure}[tp!h]
\begin{center}
\includegraphics[angle=0, width=\columnwidth]{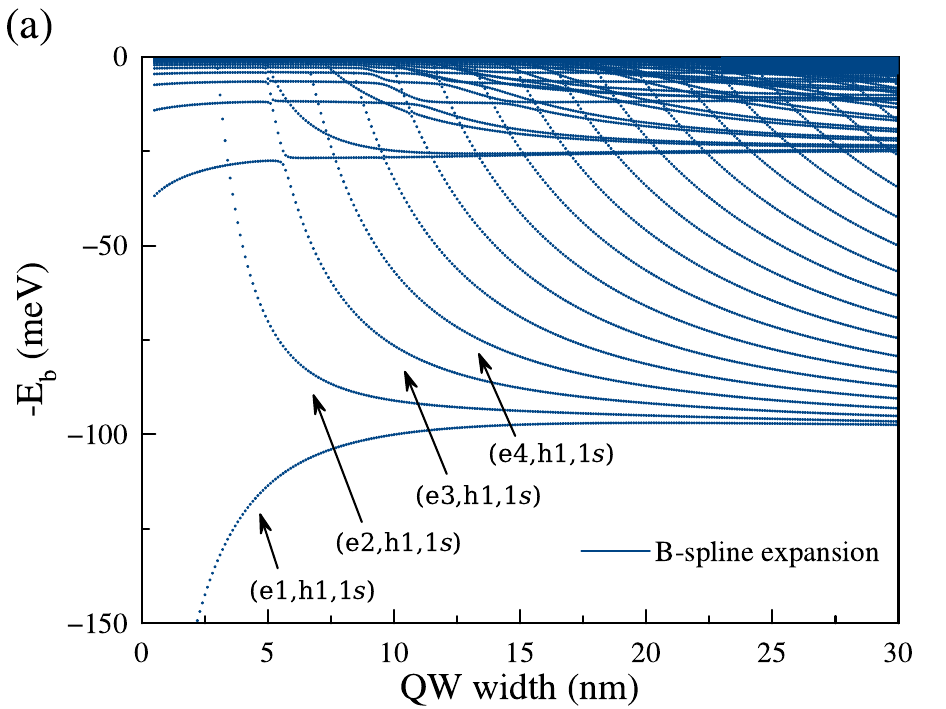}
\includegraphics[angle=0, width=\columnwidth]{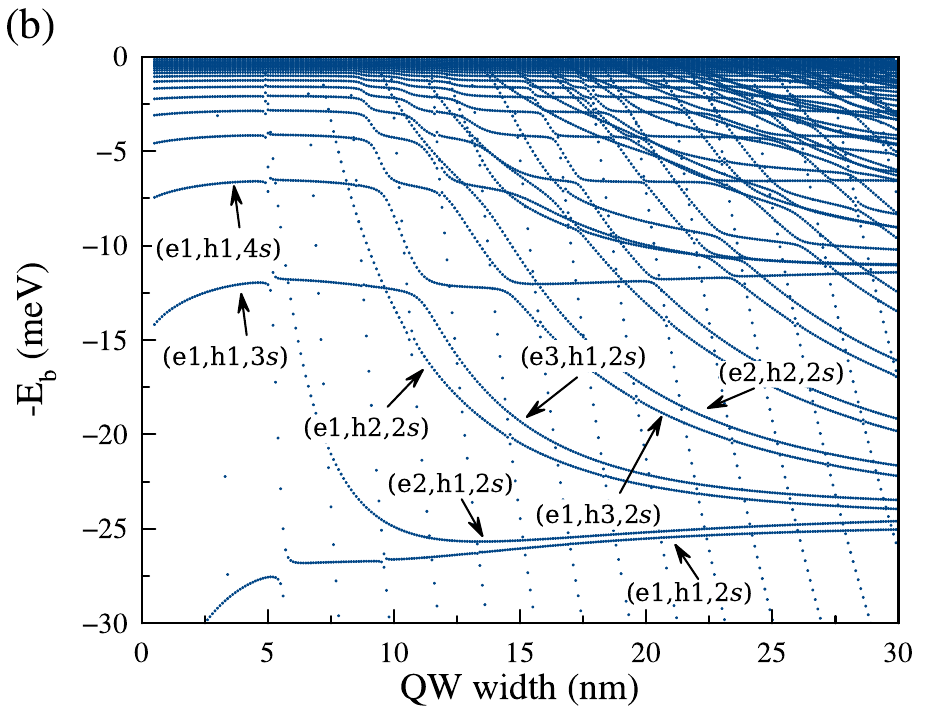}
\caption{(a) All calculated energy levels of the exciton in a Cu$_{2}$O QW as a function of the QW 
width for $m=0$. The method of the B-spline expansion of the wave function is applied. The energy levels 
are labeled as $(ei, hj, Ns)$, where $i$ and $j$ are the indices of the electron and hole quantum-
confinement states, and $N$ is the principal quantum number of the $s$-like Coulomb state. (b) Close-up 
of selected Rydberg energy levels. At this scale, the energy levels of states $(e2, h1, 1s)$ and $(e3, 
h1, 1s)$ as well as other quantum-confinement ones at panel $(a)$ look like almost vertical series 
lines.}
\label{ris1}
\end{center}
\end{figure}

The assignment of the curves is illustrated in Fig.~\ref{fig:QWExpansion}.
As a function of the QW width, the curves can be roughly divided into two groups: those which are 
approximately constant in energy over the whole range of QW widths, and those with strongly decreasing 
energies. The first group comprise states corresponding to the lowest Rydberg series, i.e., the 
exciton states below the lowest scattering threshold $E_{e1}+E_{h1}$~\cite{Belov2019}.
They are best seen by the dashed curves and the horizontal solid ones. 
These states, from bottom to top, are characterized by a triplet of quantum numbers $(e1,h1,Ns)$ with 
increasing $N=1,2,\ldots$ characterizing the energies in the Rydberg series. They accumulate around 
zero binding energy, which corresponds to the above-mentioned lowest scattering threshold.

The second group of levels comprise of those with decreasing energies as $L$ grows and which, moreover, 
converge to the energies of the first group as $L\to\infty$. The energies of the second group originate 
from the excited electron and hole quantum-confinement states $E_{ei,hj}$, where either $i>1$ or $j>1$. 
These one-particle states form the subbands in the energy spectrum~\cite{Belov2019}. Each subband has a 
scattering threshold, $E_{ei}+E_{hj}$, producing a certain branch of the continuum.
Below each scattering threshold there is a proper Rydberg series of energy levels. For small QW widths, 
the quantum confinement is strong, and the energy separation between one-particle states is large.
Therefore, for small $L$ the upper subbands lie high in the continuum, far above the lowest scattering 
threshold $E_{e1}+E_{h1}$~\cite{Belov2019}. These $eh$ states are resonant (or quasi-bound) ones.
As a result, there is only one, namely the lowest, Rydberg series of 2D Coulomb-like $eh$ bound states 
in the spectrum for small QW widths.
As $L$ increases, the strength of the confinement is gradually reduced and the energy distance between 
quantum-confinement states decreases. The upper subbands gradually decrease in energy causing a 
penetration of the electron-hole resonant states below the threshold $E_{e1}+E_{h1}$, making them bound.
As a result, with increasing $L$, more $eh$ bound states, exciton states, appear in the spectrum.
For cuprous oxide $m_{e}>m_{h}$, hence the lowest states which appear in the discrete part of the 
spectrum are those with the excited electron eigenmodes $ei$, $i=2,3,\ldots$~.

In analogy to the parabolic confinement, we also see in Fig.~\ref{fig:QWExpansion} that the alignment 
of the exciton states, obtained from the expansion for the weak confinement, along the $z$ axis 
determines their order with respect to energy. This can be seen, for example, by tracing the lowest 
states which converge to the $N=1$ or $N=2$ Rydberg energy levels. As $L$ decreases, the lowest state 
$(1X,1s)$ turns into the $(e1,h1,1s)$ state. Similar to the one-particle states in the QW, the second 
lowest state, $(2X,1s)$ is above $(1X,1s)$ due to its more spread-out wave function over the confinement 
direction.

The assignment becomes particularly complicated for the higher-lying states due to the coupling with a 
large number of confinement states. Moreover, even for the lower energy levels, the strong- and weak-confinement expansion calculations poorly match in the intermediate region of QW widths.
To reliably calculate the energies in the intermediate region between the strong and weak confinement, 
as well as in order to correctly identify the energy levels of the higher states, a precise numerical 
method is required.

Similar to the previous sections of parabolic confinement, the energy levels of the exciton in the 
rectangular QW can be calculated using the finite-difference approach~\cite{Khramtsov,Belov2019}.
However, an accurate solution of the three-dimensional Eq.~\eqref{eq:3Deq} by finite-differences is 
feasible only for a few low eigenstates.
For a precise determination of many Rydberg energies we applied the more powerful B-spline expansion 
method~\cite{deBoor}. The numerical details of this method are given in Appendix~B.

We calculated the energy levels of the exciton in Cu$_{2}$O-based QWs of various widths ranging from 
0.2~nm until 50~nm, which includes the crossover from the limits of narrow QWs, when $L \ll a_{B}$, to 
that of wide QWs, when $L \gg a_{B}$.
For better visibility, in Fig.~\ref{ris1} we show the obtained binding energies of different exciton 
states as function of the QW width for $L<30$~nm.
The binding energies are presented with respect to the lower boundary of the continuum, 
$E_{e1}+E_{h1}$, which is conventionally denoted by the zero energy level.
In Fig.~\ref{ris1}, panel $(a)$ shows all calculated energy levels, whereas panel $(b)$ presents the 
enlarged domain of the Rydberg energy levels ($Ns$, $N=2,\ldots,9$) and the excited quantum-confinement 
states to highlight the crossings and avoided crossings.

The calculated $eh$ bound states in Fig.~\ref{ris1} were subject to the same classification, 
$(ei, hj,Ns)$, based on the types of their in-plane relative and quantum-confinement motions as in 
Fig.~\ref{fig:QWExpansion}.
Such a classification is exact for narrow QWs. For wider QWs, it is only approximate.
It represents the dominant pure state  $|ei, hj, Ns \rangle$ of the exciton in a very narrow QW, 
inherent in the calculated state for a given $L$.

\begin{figure}[tp!h]
\begin{center}
\includegraphics[angle=0, width=\columnwidth]{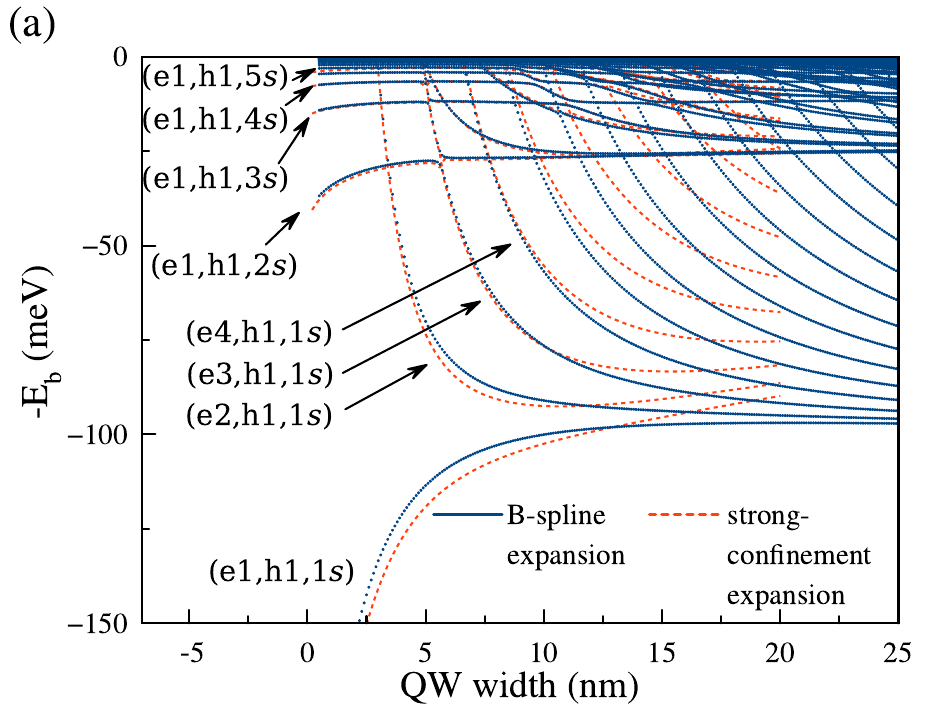}
\includegraphics[angle=0, width=\columnwidth]{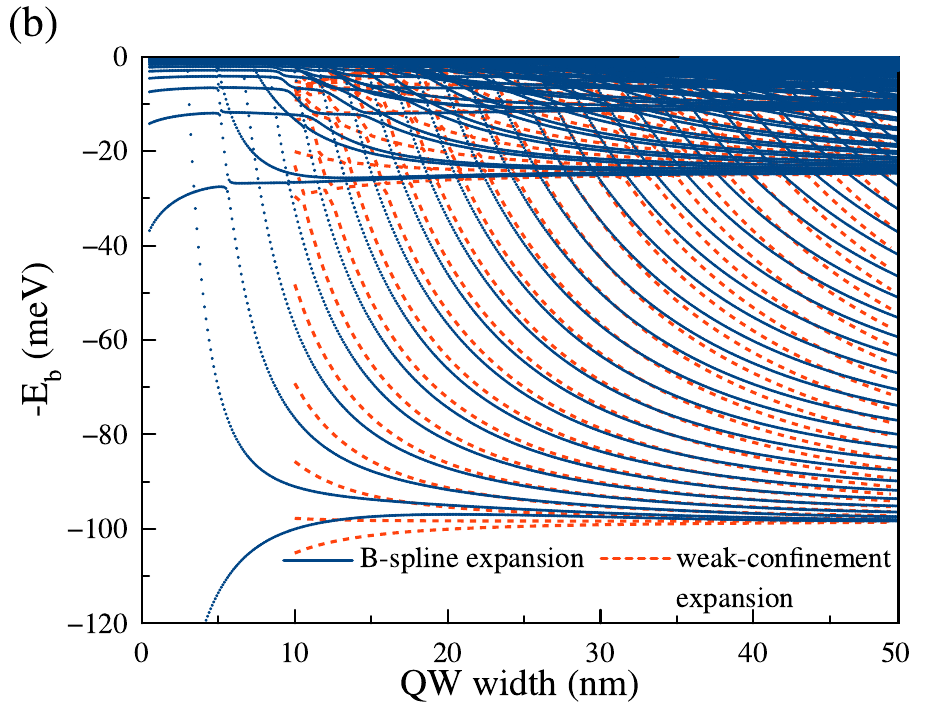}
\caption{Comparison of (a) the strong-confinement and (b) the weak-confinement expansion calculations with the data obtained from B-spline expansion. For simplicity, only $s$-like states are shown.}
\label{ris2}
\end{center}
\end{figure}

Figure~\ref{ris1}~$(b)$ shows the exciton Rydberg energy levels ($2s-9s$-like states), their evolution 
with increase of the QW width and the crossings/avoided crossings with other identified 
quantum-confinement energy levels. If there are two states of the same symmetry of the 
quantum-confinement eigenmodes, for example $(e3, h1, 1s)$ and $(e1, h1, 2s)$, then the corresponding 
energy levels show an avoided crossing. If there are states of different symmetry, then their energies 
cross, see for example the states $(e1, h1, 2s)$ and $(e2, h1, 1s)$.

A comparison of the B-spline numerical results with limiting model data is shown in Fig.~\ref{ris2}.
Panel $(a)$ demonstrates the data obtained using the B-spline expansion together with the results of 
the strong-confinement expansion calculation.
Panel $(b)$, in turn, shows the numerical data confronted to the energies from the weak-confinement 
approximation of the exciton wave function.
One can see that the strong- and weak-confinement approximations work well in the corresponding limits 
of narrow and wide QWs, respectively.
The strong-confinement approximation allows one to precisely determine the series of Rydberg energy 
levels for narrow QWs, i.e for the 2D Coulomb potential $-\rho^{-1}$.
For wider QWs, when the $eh$ motion is no longer two-dimensional, this approximation is inaccurate.
Instead, the weak-confinement approximation is precise. One can see the improving agreement between 
lowest quantum-confinement states as $L$ grows. The upper quantum-confinement energy levels show less 
precise correspondence due to inaccuracy in the B-spline approximation of highly oscillating 
wave functions of the upper QW confined states. Increase of the size of the B-spline basis gives an 
improvement in the accuracy.

\begin{figure}[t!]
\begin{center}
\includegraphics[angle=0, width=\columnwidth]{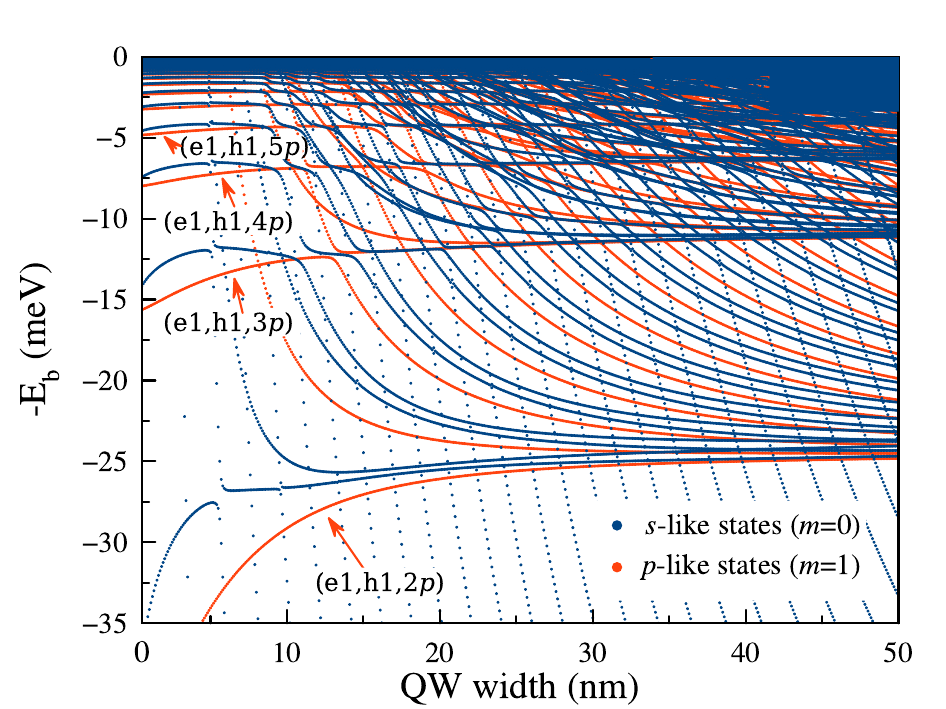}
\caption{Calculated $p$-like ($m=1$) energy levels of the exciton in a Cu$_{2}$O QW compared to the $s$-like ones ($m=0$). The method of the B-spline expansion of the wave function is applied. The energy levels are labeled as $(ei, hj, Np)$, where $i$ and $j$ are the indices of the electron and hole quantum-confinement states, and $N$ is the principal quantum number.}
\label{risM1}
\end{center}
\end{figure}

In order to compare our two-band model and the energies it produces with the experimental 
measurements~\cite{NakaPRL,Naka2018}, we calculated energy levels of the exciton states with $|m|=1$ as 
a function of the QW width, see Fig.~\ref{risM1}.
In contrast to the case of cylindrically symmetrical states, with $m=0$, which we call $s$-like states, 
those with $|m|=1$ can be attributed to $p$-like states.
The numerical method allowed us to obtained the energies up to at least the $10p$ Rydberg level. 
Similar to the $s$-like energies, among $p$-like states there are many quantum-confinement energy levels as 
well as their crossings and avoided crossings. However, despite the general similarity of the 
dependencies of $s$-like and $p$-like energy levels on the QW width, the detailed comparison reveals 
noticeable discrepancies of the states of two different kinds. These differences are especially 
pronounced for energy levels of the lowest quantum-confinement subband, $(e1,h1,Ns)$ and $(e1,h1,Np)$, 
with $N=2,\ldots, 5$ for the QW width $L<15$~nm.
In complete analogy to the parabolic confinement, the $s$- and $p$-like states degenerate only in the 
limiting cases of strong and weak confinement.
For the intermediate region of $0<L< \infty$ the energies of $p$-like states are lower than the 
corresponding $s$-like ones.
These discrepancies originate from the QW barriers, which break the exact spherical symmetry of the 
hydrogen-like exciton, leaving behind only a cylindrical symmetry. This leads to different energies of 
the states with different absolute values of the magnetic quantum numbers (and also with different 
values of orbital ones) for the same principal quantum numbers.

Our energy levels of the $p$-like exciton states can be compared to the photoluminescence spectra 
measured for different QW widths, see Ref.~\cite{Naka2018}.
The peak positions of $2p$- , $3p$-, and $4p$-like exciton states of the measured spectra well agree 
with the corresponding calculated energy levels. In particular, one can see in Ref.~\cite{Naka2018}, 
Fig. 2~(a), that the energy differences between $3p$- and $2p$-like as well as between $4p$- and 
$3p$-like states are 14~meV and 5~meV, respectively. Very similar energy differences are observed in 
Fig.~\ref{risM1} for QW width $L\sim 50$~nm, which means a good agreement between experiment and theory.

\begin{figure}[tp!h]
\begin{center}
\includegraphics[angle=0, width=\columnwidth]{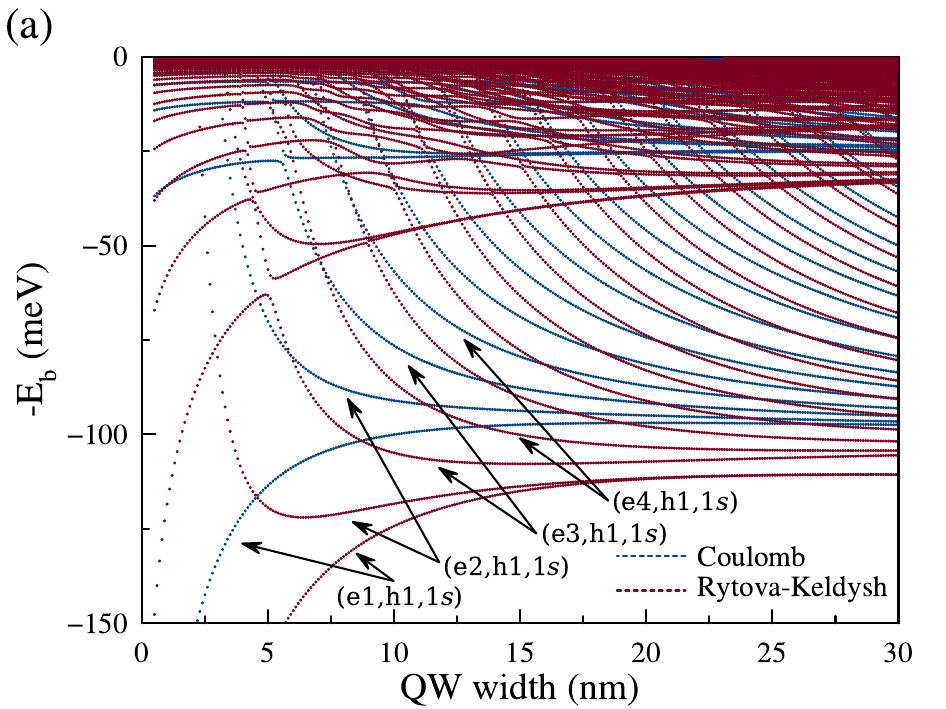}
\includegraphics[angle=0, width=\columnwidth]{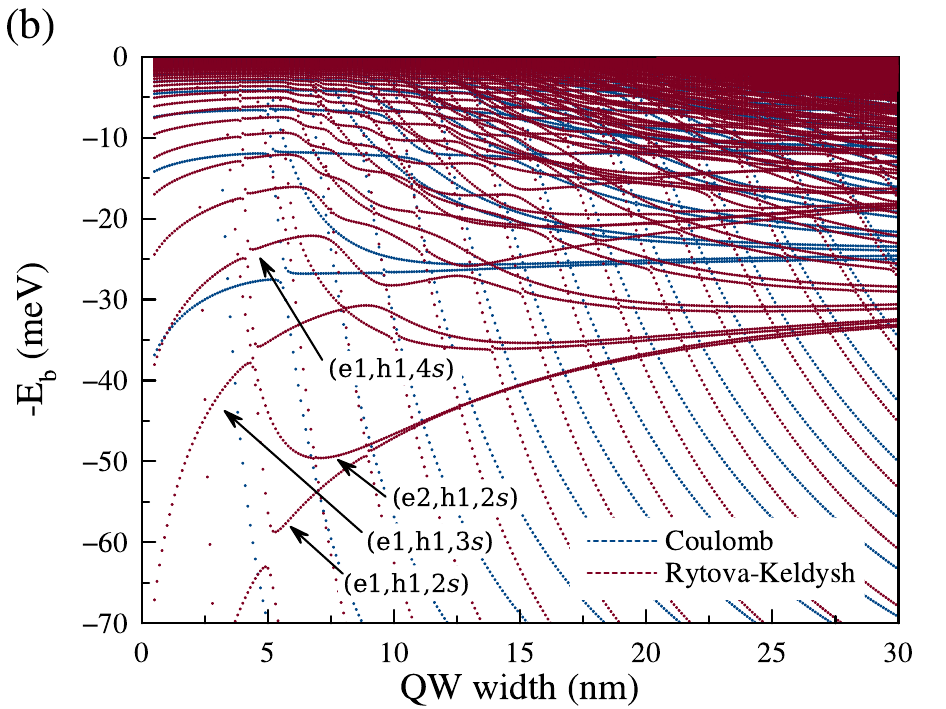}
\caption{(a) The comparison of energy levels of the exciton in a Cu$_{2}$O QW as a function of the QW 
width for $m=0$. The energy levels are obtained with the Coulomb potential and the Rytova-Keldysh potential (with the dielectric constant in the substrate $\epsilon_{b}=3$). The energy levels 
are labeled as $(ei, hj, Ns)$, where $i$ and $j$ are the indices of the electron and hole quantum-
confinement states, and $N$ is the principal quantum number of the $s$-like Coulomb state. (b) Close-up 
of selected Rydberg energy levels.}
\label{CompareRKCoulomb}
\end{center}
\end{figure}

\section{Electrostatic effect of the barrier material on the electron-hole interaction}
\label{sec:RK}
The quantum confinement is the main effect determining the energy levels. However, it is 
not the only effect. Facets of the heterojunction of the cuprous oxide crystal embedded into 
air or in a substrate (sapphire or quartz) are characterized by different dielectric constants.
In order to describe an experimental setup as in Ref.~\cite{Naka2018} in which a thin film of 
Cu$_{2}$O is sandwiched by a substrate material, we consider an effect of the dielectric contrast 
in the film and in the substrate to the energy levels of the Rydberg exciton.
The heterostructure with different dielectric permittivities leads to different rates of charge 
screening and thus to a distortion of the Coulomb interaction between charge carriers.
This is especially important for narrow structures in which the electrostatic field is mainly concentrated in the substrate,
but not in the film~\cite{Chernikov,Kez2017}.
The electron-hole interactions in a heterostructure with dielectric contrast are rigorously 
described by the Rytova-Keldysh potential~\cite{Rytova,Keldysh}. It defines the infinite series of 
the Coulomb-like interaction potentials between the given charge and the fictional image charges.

For an electron and a hole in a structure with the dielectric constant $\epsilon_{\mathrm{QW}}$ 
in the QW and $\epsilon_{\mathrm{b}}$ in the barrier, the Rytova-Keldysh potential is given in 
Fourier space by
\begin{align}
\label{RKorig}
    \nonumber & V(\vec{\rho},z_{e},z_{h}) = -\frac{4\pi e^{2}}{\epsilon_{\mathrm{QW}}}
    \int \frac{d^{2}k}{(2\pi)^2} e^{i \mathbf{k} \vec{\rho}}\nonumber \\[1ex] &\times
    \frac{\cosh{\left[k(L/2-z_{e})+\eta \right]} \cosh{\left[k(L/2+z_{h})+\eta \right]}}{k \sinh{\left[kL+2\eta \right]}},
\end{align}
where $\eta=\frac{1}{2} \ln{\frac{\epsilon_{\mathrm{QW}}+\epsilon_{\mathrm{b}}}
{\epsilon_{\mathrm{QW}}-\epsilon_{\mathrm{b}}}}$. When using a series expansion for the 
denominator, i.e.
\begin{equation}
  \frac{1}{\sinh x} = 2\sum_{n=0}^\infty\mathrm{e}^{-(2n+1)x}
\end{equation}
for $x>0$, the Fourier integrals can be solved analytically resulting
in the potential
\begin{align}
\label{eq:RKexpansion}
  V(\rho,z_{e},z_{h}) = -\frac{{e}^2}{\epsilon_{\mathrm{QW}}}
  \sum_{n=-\infty}^{+\infty}\frac{\gamma^{|n|}}{\sqrt{\rho^2+(z_{{e}}-z_{{h}}^{(n)})^2}} \, ,
\end{align}
with
$\gamma=\frac{\epsilon_{\mathrm{QW}}-\epsilon_{\mathrm{b}}}{\epsilon_{\mathrm{QW}}+\epsilon_{\mathrm{b}}}$.
Here, 
$z_{{h}}^{(n)}=(-1)^nL-z_{{h}}^{(n-1)}$ with $z_{{h}}^{(0)}=z_{{h}}$
are the $z$ positions of introduced image charges of the hole after $|n|$ alternating mirror
reflections at the two barrier surfaces at $z=\pm L/2$ in complementary order for $n\gtrless 0$.
Note that the image charges are located on a line perpendicular to the
barrier planes of the QW and decrease exponentially with the number $|n|$
of the reflections.

In the 2D limit, the potential~\eqref{RKorig} is reduced to the effective potential, given by the 
difference of the Struve function $S(x)$ and the Bessel function of the second kind $Y_{0}(x)$ as
$$
V_{\text{eff}}(\rho) = -\frac{\pi e^{2}}{L \epsilon_{\mathrm{QW}}} \left[ S(\rho/\rho_{0}) - Y_{0}(\rho/\rho_{0}) \right].
$$
Here $\rho_{0}=L\epsilon_{\mathrm{QW}}/(2\epsilon_{\mathrm{b}})$ is the parameter defining the scale of the interactions.
This potential has a singular behavior $V_{\text{eff}}(\rho)\sim -\ln{\rho}$ as $\rho \to 0$ and 
the Coulomb-like asymptotic form $V_{\text{eff}}(\rho)\sim -\rho^{-1}$ as $\rho \to \infty$~\cite{Cudazzo,Juan}.

It is worth noting that the dimensional reduction also introduces a similar logarithmic 
divergence of the effective 2D potential of the pure Coulomb interaction~\cite{Florez}.
The point is that the averaging of the Coulomb potential over the quantum confinement wave 
functions results in the effective 2D interactions in a such way that it behaves as 
$\sim -\ln{\rho}$ as the 2D radius $\rho\to 0$. 
The direct B-spline-expansion solution of the 3D Schr\"{o}dinger equation is equivalent to the 
method of the solution by averaging of the problem over the $z$ direction and then solving the 
effective equation for the in-plane motion (over $\rho$) with the potential~\eqref{Veff}. Thus, 
during a crossover to the 2D geometry, our effective 2D potential simulates a logarithmic 
divergence as $\rho\to 0$. 

As a result, the effect of the dielectric contrast at the heterojunction does not introduce new 
singularities of the potential. Therefore, the general structure of the Rydberg and quantum-
confinement energy levels should still hold and only the details change.
The strongest effect of the Rytova-Keldysh potential occurs for narrow QWs. As the strength of the 
confinement depends on the quantum number of a given state, the Rytova-Keldysh potential can also 
noticeably change the Rydberg states of the confined exciton for $L\gg a_{B}$.

To study this issue, we calculated the energy spectrum of the $eh$ bound states with the 
interactions described by the Rytova-Keldysh potential.
We used the dielectric constant $\epsilon_{\mathrm{QW}}=7.5$ in QW and $\epsilon_{\mathrm{b}}=3$ in 
the barrier.
The latter is a typical value for a sapphire or quartz substrate~\cite{Polyanskiy}.
With these values the image charge $e\gamma^{|n|}$ in Eq.~\eqref{eq:RKexpansion} is reduced by about three orders of 
magnitude after eight mirror reflections, which guarantees the fast convergence of the potential.
A comparison of the results with the Rytova-Keldysh and the Coulomb potential is shown in 
Fig.~\ref{CompareRKCoulomb}.
One can see that, generally, up to an additive constant, the structure of energy levels is similar 
for both potentials.
Nevertheless, one observes that the dielectric contrast with 
$\epsilon_{\mathrm{b}}<\epsilon_{\mathrm{QW}}$ shifts the energies downward, hence the ground and 
excited states become more localized, and eventually more bound states appear in the spectrum.
This means that, due to the lower $\epsilon_{\mathrm{b}}$ in the substrate, the $eh$ interaction 
becomes stronger. The penetration of the electrostatic field in the barrier material is increased 
due to the factor $1/\epsilon_{\mathrm{b}}$, compared to $1/\epsilon_{\mathrm{QW}}$ in the QW.
This also leads to larger avoided crossings.
Moreover, the region where the avoided crossings take place is naturally a transition region from 
weak to strong confinement. For the Rytova-Keldysh potential, this region, as a whole, is shifted 
to the lower QW widths. This is again the result of the increased $eh$ interaction in the barrier 
material, compared to the case when $\epsilon_{\mathrm{b}}=\epsilon_{\mathrm{QW}}$.
The upper states with, for example, quantum numbers $(e1,h1,4s)$ and Rydberg exciton states with 
higher principal quantum number in Fig.~\ref{CompareRKCoulomb}(b) are clearly in the transition 
region or in the strong confinement region for $L$ up to 30 nm.

One can also see from the figure that some states degenerate more rapidly as $L \to \infty$ with 
the Rytova-Keldysh interactions than with the Coulomb potential. This takes place, for example, for 
the $(e1,h1,1s)$ and $(e2,h1,1s)$ states. This is a direct consequence of the shifted transition 
region. The degenerate energy levels at smaller QW widths mean that, for the Rytova-Keldysh 
potential, the relatively weaker confinement appears at smaller $L$, due to the relative increase 
of the $eh$ interactions (when $\epsilon_{\mathrm{b}}<\epsilon_{\mathrm{QW}}$), than in the case of 
the pure Coulomb potential.

A comparison of the Rytova-Keldysh results with the photoluminescence data given in 
Ref.~\cite{Naka2018} shows reproducing the same energy differences as the Coulomb interactions do. 
The point is that both the calculated $p$-like energy levels for the weak confinement 
asymptotically as $L\to \infty$ converge to the same values. The measurements were done for a 
thickness $L \sim 100$~nm and larger for which our calculations with both the potentials produce 
the same energy differences.

\section{Effects of the crystal environment}
\label{sec:valence}
So far, we used the hydrogen-like two-band model to describe the kinetic energies of the 
electron and hole. This model adequately reproduces the general structure of Rydberg energy levels, 
but does not show all the features of the spectrum such as the fine-structure splitting.
If a sufficient accuracy to allow for a quantitative comparison with experiment is desired, the 
complex valence-band structure and other central-cell corrections for small electron-hole 
separations~\cite{Joerg2} have to be considered. In this section, we briefly discuss how these 
improvements affect the presented spectra.

A more sophisticated model for the energy dispersions of the electron and hole, that is
able to properly describe the non-parabolicity of the valence bands, is given by the Suzuki-Hensel 
Hamiltonian, based on the Luttinger-Kohn 
model~\cite{Luttinger55,Luttinger56,SuzukiHensel74,Schweiner16b}. In this description, the 
treatment of the kinetic energy of the electron and hole requires the introduction of additional 
spin degrees of freedom, viz., the quasispin and the hole spin. With these quantities, all symmetry-compatible terms up to quadratic order in the hole momentum are included, leading to a much more 
complicated total Hamiltonian and an increase of the dimension of the Hilbert space by a factor of 
six. As a result, the computation of the fine-structure splitting due to the non-parabolic 
dispersion of the hole is nontrivial for highly-excited Rydberg states in the transition 
region from weak to strong confinement.

Further complication of the model includes the central-cell corrections. They mainly affect 
the exciton states with principal quantum number $N=1$, marked as $1s$ in Figs.~\ref{fig:QWExpansion}-\ref{ris2}, and \ref{CompareRKCoulomb}. These 
states have the smallest spatial extension, and therefore most quickly reach the weak confinement 
region with increasing QW width.
Most corrections will result in a simple shift of the energy, except for the exchange interaction, 
which causes a splitting of $s$-like states into the threefold degenerate $\Gamma_5^+$ orthoexcitons and 
a $\Gamma_2^+$ paraexciton~\cite{Joerg2,Rommel2021,koster63}. The bulk $1s$ exciton is, in 
particular, split by about $12$~meV~\cite{Joerg2}. A similar fine structure splitting is expected 
for the different $1s$-like states originating from the quantization in QW, i.e., there are two 
sequences of $1s$-like states with symmetries $\Gamma_2^+$ and $\Gamma_5^+$, respectively.
For each symmetry, the spectra should be very similar (up to an energy offset) to the Rydberg 
series obtained with the two-band model, but for symmetry reasons in the QW, an additional 
splitting of the $\Gamma_5^+$ excitons into a one-dimensional and two-dimensional subspace is 
expected~\cite{koster63}.

Additionally, in the QW the translational symmetry is broken along the $z$ direction, i.e., 
perpendicular to the QW plane. As a result, the associated center-of-mass momentum is no longer a 
conserved quantity. Thus, the relative and center-of-mass coordinates cannot be introduced for all 
three spatial directions as in the bulk, but only in the QW plane.
The rotational symmetry around the $z$ axis is also broken by the crystal. Therefore, the 
computation of the fine-structure splitting requires the solution of the Schr\"odinger equation 
with four degrees of freedom in coordinate space, in addition to the spin degrees of freedom 
related to the quasispin and hole spin. The derivation and implementation of the full Hamiltonian 
by using a complete basis set with, e.g., Sturmian or B-spline functions, is a highly nontrivial 
task. Furthermore, due to the increased size of the Hilbert space, the numerical effort for 
diagonalizing the Hamiltonian will grow drastically compared to the hydrogen-like two-band model 
for excitons in QWs.

\section{Conclusions}
\label{sec:results}

In this work, we have computed the energy levels of Rydberg excitons in cuprous oxide QWs of various 
thicknesses, ranging from weak confinement, where the effect of QW barriers can be treated as a 
small perturbation, to strong confinement, where the Coulomb potential degenerates into its two-dimensional counterpart.
The limiting cases of weak and strong confinement can be treated by Coulomb-Sturmian and 
adiabatic expansion methods, respectively. However, the crossover from weak to strong confinement 
across ten orders of magnitude of the confinement strength can only be accurately studied by numerical 
methods, in our case using a finite-difference approximation or B-spline expansion methods.

We obtained the Rydberg series of energy levels in case of parabolic confinement, observed their 
dependence on the strength of the confinement, the crossings and avoided crossings of energies. The 
parabolic confinement has the distinct advantage that the variables in the Schr\"odinger equation are 
separable, and the motion of the exciton as a whole in the QW is independent of the relative $eh$ 
motion. We observed the evolution of the energy spectrum during a crossover from weak to strong 
confinement, i.e., from a wide QW to a very narrow one. We observed that, due to confinement, the states 
that are mainly aligned along the QW plane, remain in the spectrum of bound states.

For rectangular confinement, we calculated the energy levels of excitons in Cu$_{2}$O-based quantum
wells as a function of the QW width. In this case the center-of-mass motion along the confining 
direction cannot be separated and one has to solve the full three-dimensional problem.
The B-spline expansion of the exciton wavefunction allowed us to calculate the 
quantum-confinement and the Rydberg energy levels for a whole range of QW widths, including the 
crossover from strong to weak confinement regimes.
We observed crossings and avoided crossings of energy levels and reproduced their well-known behavior 
in the limiting cases.
We showed that the limiting models are precise in the corresponding confinement regimes, however they 
are unable to adequately describe the energy levels in the intermediate range of QW widths.
Nonetheless, these models provide us with an insight as to how to classify the numerically obtained 
energy levels. We properly identified and classified several energy levels as well as explained their 
crossings and avoided crossings based on the symmetry properties of their quantum-confinement 
eigenmodes.

We additionally considered the effect of the dielectric contrast between the QW and the 
barrier. The smaller dielectric permittivity in the substrate causes an increase of the $eh$ 
interaction and thus leads to larger binding energies of the states. Although our investigations 
using the Rytova-Keldysh potential are more related to the real experimental setup, other issues 
raised by the dielectric contrast, for example a breakup of the exact parity due to the dielectric 
asymmetry, have to be studied in the future.

In this paper, we have used a simple two-band model to describe the kinetic energy of the 
electron and hole. This means that effects of the complex valence band structure on Rydberg 
excitons in QWs have been neglected. The accurate consideration of these effects, as outlined in 
Sec.~\ref{sec:valence}, will be important for future detailed line-by-line comparisons between 
experiment and theory.
Nonetheless, our results can be compared to the available experimental photoluminescence and 
reflectance spectra of excitons in Cu$_{2}$O-based thin films of different thicknesses. For 
example, a good agreement was obtained between the calculated results and the photoluminescence 
spectra from Ref.~\cite{Naka2018}. We believe that our data will facilitate further experimental 
studies and interpretation of the reflectance and photoluminescence spectra of excitons in thin 
films of cuprous oxide.

\acknowledgements
The authors are grateful to anonymous referees and J. C. del Valle for constructive comments which significantly improved the exposition.
This work was financially supported by the Deutsche Forschungsgemeinschaft through SPP 1929 GiRyd, projects SCHE~612/4-2, MA~1639/16-1, and GI~269/14-2.

\section*{Appendix}
In the Appendix, we briefly outline the numerical methods which are used to calculate the energy levels for arbitrary strength of the confinement (i.e.\ for arbitrary QW width).

\subsection{Finite-difference approximation}
\label{AppA}

On an equidistant grid with step size $h=\rho_{i+1}-\rho_{i}$, the finite-difference approximation of 
the second-order derivatives of Eq.~\eqref{exParabolic} is given as
$$
\frac{\partial^{2}\Phi(z,\rho)}{\partial\rho^{2}} = \frac{\Phi(z,\rho_{i-1})-2 \Phi(z,\rho_{i})+\Phi(z,\rho_{i+1})}{h^{2}}+O(h^{2}).
$$
The first-order derivative is approximated by
$$
\frac{\partial\Phi(z,\rho)}{\partial\rho} = \frac{\Phi(z,\rho_{i+1})-\Phi(z,\rho_{i-1})}{2h}+O(h^{2}).
$$
Using these formulas in Eq.~\eqref{exParabolic} leads to the eigenvalue problem with a five-diagonal 
matrix~\cite{arrowhead}. Several lowest eigenvalues of this matrix are calculated using ARPACK package~\cite{ARPACK}.
As a result, the energies of the $eh$ bound states for different strengths of the parabolic confinement are obtained.

\subsection{B-spline basis expansion}
\label{AppB}

The B-splines of higher orders are a more effective tool for discretizing and solving the partial 
differential equations than the finite-difference approximation. The unknown function is expanded over 
a basis of B-splines $B^{k}_{i}(x)$, $i=1,\ldots,n$, which are piecewise polynomials of degree $k-1$.
Given the predefined series of the 
service nodes $t_{i}$, each B-spline $B^{k}_{i}(x)$ of order $k$ is defined on the interval $[t_{i},t_{i+k}]$.
Values of the B-splines $B^{k}_{i}(x)$ and their derivatives at a given point $x$ can be calculated by 
recursion formulas~\cite{Bachau}
$$
B^{k}_{i}(x)=\frac{x-t_{i}}{t_{i+k-1}-t_{i}} B^{k-1}_{i}(x)+
\frac{t_{i+k}-x}{t_{i+k}-t_{i+1}} B^{k-1}_{i+1}(x),
$$
$$
\frac{d B^{k}_{i}(x)}{dx} = \frac{k-1}{t_{i+k-1}-t_{i}} B^{k-1}_{i}(x)-
\frac{k-1}{t_{i+k}-t_{i+1}} B^{k-1}_{i+1}(x).
$$
The expression for the second derivative can be derived from the above two equations.

Three important characteristics of B-splines should be highlighted.
First, a B-spline of order $k$ on an equidistant grid approximates an analytical function with accuracy 
of about $h^{k}$, where $h$ is the step size of the grid.
Thus, higher-order B-splines give an accurate solution even for a relatively small number of nodes, 
which is particularly important for multi-dimensional problems.
Second, one can choose nodes non-equidistantly and add service nodes at the boundaries in such a way as to 
have some B-splines equal to 1 at the boundaries, while all other B-splines are exactly zero there.
Then, for example, zero boundary conditions can be easily implemented by removing the B-splines which 
are nonzero at the boundaries.
Third, the B-splines are nonorthogonal functions, and thus the problem turns into a generalized 
eigenvalue problem.
However, the B-spline functions have minimal support, i.e., each B-spline
vanishes, $B^{k}_{i}(x) = 0$, for $x \notin [t_{i},t_{i+k}]$, which significantly reduces the number of 
integrations to calculate matrix elements. This leads to a sparse structure of the matrices of the 
generalized eigenvalue problem.

When applying B-splines to represent the Hamiltonian~\eqref{eq:3Deq} of the exciton in a QW, the following should be noted:
although one applies Dirichlet boundary conditions as $\rho \to \infty$, and one can restrict the 
calculation domain by setting a large cutoff $\rho=\rho_{\max}$, the boundary conditions at $\rho=0$ are less obvious. The wave function $\psi$ should be finite, but not necessarily be zero at $\rho=0$.
To construct a zero boundary condition at $\rho=0$, we use the substitution $\psi=\chi/\sqrt{\rho}$.
As a result, the B-spline expansion of $\chi$ is employed to the Hamiltonian
\begin{align}
  \label{eq:Hamiltonian_substitution}
	H &= - \frac{\hbar^2}{2\mu}\left(\frac{\partial^2}{\partial\rho^2}
	-\frac{m^2-1/4}{\rho^2}\right)
	- \frac{\hbar^2}{2m_{e}}\frac{\partial^2}{\partial z_{e}^2}
	- \frac{\hbar^2}{2m_{h}}\frac{\partial^2}{\partial z_{h}^2} \nonumber\\
	&-
	\frac{e^2}{\epsilon\sqrt{\rho^2+(z_{e}-z_{h})^2}}.
\end{align}
The Coulomb interaction in this equation can be easily replaced by the Rytova-Keldysh potential~\eqref{eq:RKexpansion}.
The zero boundary conditions over z are defined at $z_{e,h}=\pm L/2$.
The constructed boundary value problem with zero boundary conditions allows one to accurately
approximate the spectrum of bound states below the scattering threshold.

It is also worth noting that the fundamental solution of the in-plane radial part of 
Eq.~\eqref{eq:Hamiltonian_substitution} is a linear combination of Bessel functions. One of these 
functions diverges as $\rho\to 0$. To avoid the divergence for magnetic quantum number $m=0$, the zero boundary condition at $\rho\to 0$
was shifted to $\rho= -\varepsilon<0$, where $\varepsilon$ is by three orders of magnitude smaller than 
the smallest step of the grid.
For $|m|>0$ the divergence was not observed and no shifts were applied.

The B-spline basis expansion reads as
$$
\chi(\rho,z_{e},z_{h})=\sum_{ijk} c_{ijk}\, B^{k}_{i}(\rho) B^{k}_{j}(z_{e}) B^{k}_{k}(z_{h}),
$$
where the coefficients $c_{ijk}$ form the eigenvectors of the generalized eigenvalue problem.
In our calculations, we used B-splines of order $k=5$ with
equidistant nodes along the $z$ direction and non-equidistant nodes along the $\rho$ axis.
For the non-equidistant grid, the interval between nodes scales cubically with its number.
We used 30 nodes for the $\rho$-direction and 22 nodes for each of two $z$-coordinates.
The matrix elements were calculated numerically by application of a 15 point Gauss-Kronrod formula~\cite{Kahaner89}. The resulted generalized eigenvalue problem was solved by ARPACK routines~\cite{ARPACK}.


\bibliography{paper}

\end{document}